\documentclass[11pt,floatfix,preprint,showpacs]{revtex4}
\usepackage{epsfig}
\usepackage{graphicx}
\usepackage{dcolumn}
\usepackage{bm}
\usepackage{amsmath}
\usepackage{amssymb}
\usepackage{amsfonts}

\def\e{\kern+.6ex\lower.42ex\hbox{$\scriptstyle \iota$}\kern-1.20ex e}
\begin{document}

\title{3N Scattering in a Three-Dimensional Operator Formulation}

\author{W.\ Gl\"ockle$^1$}

\author{I. Fachruddin$^2$}

\author{Ch. Elster$^3$}

\author{J.~Golak$^4$}

\author{R.~Skibi\'nski$^4$}

\author{H.~Wita{\l}a$^4$}

\affiliation{$^1$Institut f\"ur theoretische Physik II,
Ruhr-Universit\"at Bochum, D-44780 Bochum, Germany}

\affiliation{$^2$Departemen Fisika Universitas Indonesia, 
Depok 16424, Indonesia}

\affiliation{$^3$Institute of Nuclear and Particle Physics, 
Department of Physics and Astronomy, Ohio University, Athens, OH 45701, USA}

\affiliation{$^4$M. Smoluchowski Institute of Physics, Jagiellonian
University, PL-30059 Krak\'ow, Poland}

\date{\today}
\begin{abstract}
A recently developed formulation for a direct treatment of the 
equations for two- and three-nucleon bound states as set of coupled
equations of scalar functions depending only on vector momenta is
extended to three-nucleon scattering. 
Starting from the spin-momentum dependence occurring as scalar
products in two- and three-nucleon 
forces together with other scalar functions, we present the Faddeev multiple
scattering series in which order by order the spin-degrees can be 
treated analytically leading to
3D integrations over scalar functions depending on momentum vectors only.
 Such formulation is especially important in view of awaiting 
 extension of 3N Faddeev calculations to projectile energies above 
the pion production threshold
 and applications of chiral perturbation theory 3N forces, which are to be most
 efficiently treated directly in such three-dimensional formulation 
without having to expand these forces into  a partial wave basis. 
\end{abstract}
\maketitle \setcounter{page}{1}

\section{Introduction}
\label{sec1}

The three-nucleon (3N) system is the first nontrivial case to learn 
about the action
 of nucleon-nucleon (NN) and three-nucleon (3N) forces in bound states and
scattering observables. Below about 200~MeV laboratory projectile energy 3N
scattering 
can be well treated in a momentum space representation based on
the Faddeev equations in a
partial wave representation~\cite{report} to calculate
 elastic nucleon-deuteron (Nd) scattering as well as breakup
 processes. A different approach using a coordinate
space representation and based on hyperspherical harmonics 
 is equally precise for elastic Nd scattering~\cite{pisa}. 

 Over many years a rich set of Nd data has been
 accumulated~\cite{report,data,seki,stephan}, 
which is not only
a very valuable source of information about the spin- and momentum-dependence 
of nuclear forces but also about 
the reaction mechanism of multiple rescattering processes. The nuclear  forces
under consideration today are on the one hand the
so-called (semi)-phenomenological high precision
forces~\cite{av18,cdb,nijm}, 
 describing the NN data up to the pion production threshold perfectly well.
The three-nucleon forces on a semi-phenomenological level are much less 
developed~\cite{3NF} 
 and are not constructed in
a consistent manner with respect to corresponding NN forces. Nevertheless 
those NN and 3N forces describe 3N data
 (elastic as well as breakup cross sections and numerous spin
observables)  
often 
spectacularly  well.
However, there are exceptions where one finds
serious discrepancies  between that theoretical prediction  
and the data, especially for
some spin observables at the
higher region of that energy range~\cite{data,seki,stephan}.

In recent years effective chiral perturbation theory links nuclear
  forces to the symmetry of QCD and diagrammatically 
builds up the nuclear forces in a systematic expansion~\cite{epelbaum_progress}.
In this fashion NN, 3N and even  4N forces are at present consistently
generated  up to next-to-next-to-next leading order (N$^3$LO). 
Applications of these forces in the 
 few-nucleon sector~\cite{application} and for light nuclei~\cite{ncsm} 
are quite successful.
 However, the effective theory is limited in the energy regime
it can be applied to, which is related to the 
smallness of the parameter
underlying that expansion. A typical upper limit for the
applicability of those chiral forces is 100-150~MeV
nucleon laboratory projectile energy. 

Considering the nucleon projectile energies above the pion production
threshold and even going into the GeV region, 
no similar systematic approach 
to nuclear forces is yet
available. However, this energy region imposes challenging questions
not only about the nuclear force, but also about the underlying 
reaction mechanism.
One should expect an
increased importance of 3N
forces , the influence of baryon resonances, meson production, to name a few. 
Another challenging
question is an investigation of 
the limit, where hadronic degrees of freedom apply and where
subnuclear 
degrees of freedom must be explicitly considered.
In order to enter this energy region in the 3N sector 
various challenges have to be
overcome. On the technical side the standard partial-wave decomposition
 (PWD) has to be given up due to the strongly increasing
number of partial wave states that need to be summed for a
converged result. To face this challenge the direct use of momentum vectors, 
i.e. a 3-dimensional (3D) formulation of the problem turned
out to be a promising path. This has been documented in various
studies for three-boson scattering carried
out in a Faddeev scheme in momentum space~\cite{Liu:2004tv}.
In addition, the high energy region requires that Galilean invariance has to be
replaced by Poincar\'e invariance. For the case of three-boson scattering
this has successfully been achieved~\cite{Lin:2008sy,Lin:2007kg}, where the
Poincar\'e invariant Faddeev equations have been solved for projectile energies
in the GeV regime.
As already said the enormous challenge will
be to develop the underlying dynamics.
 
It turns out that Poincar\'e invariant formulation of the 3N system is 
already important
  at quite low laboratory energy. Relativistic effects
 are discernable in some regions of the breakup phase-space starting
 at the energy of the incoming nucleon about $65$~MeV~\cite{witrel1}.
 For
 Nd elastic scattering  vector analyzing power they contribute to the
 famous analyzing power puzzle at energies around
 $10$~MeV~\cite{witrel2}.

In this paper we shall focus on the question how to incorporate spin and 
isospin degrees of freedom into the bosonic three-body calculations so that
the successful approach of Ref.~\cite{Liu:2004tv} is applicable to
nucleons.
The aim is to reduce the formulation with spin/isospin degrees
of freedom to scalar,  spin independent 
functions of vector momenta in the same spirit as 
already presented for the 2N and 3N bound states in Ref.~\cite{2N3N}.
 The idea is to use the original structure of the NN forces consisting of scalar
operators in  spin- and momentum-space and 
 scalar functions which only depend on momenta. This
carries over to the NN  t-operator which is a central building block
in the Faddeev scheme. In addition, 3N forces appear naturally 
in this formulation.

One form of the Faddeev equations for 3N scattering is based on the
multiple scattering series for the
breakup process which can be  summed into a Faddeev integral equation
for a transition operator $ T|\Phi\rangle$
 in our standard notation~\cite{actapolonica}.  Here $|\Phi\rangle$  is
 the 
initial product state of a deuteron and a momentum
eigenstate of the projectile nucleon. Analogous to the NN
t-operator there will be an operator form
for three-nucleon transition operator $T|\Phi\rangle$. 
However the number of scalar spin-momentum operators
will be enormously high, which makes it
not advisable to rewrite the Faddeev equation for $T|\Phi\rangle$  into a
coupled 
set of equations for the
accompanying scalar momentum  dependent functions. The experience with
high 
energy  three-boson scattering~\cite{Liu:2004tv} suggests
that it is promising to  generate instead the multiple scattering
series,  which will automatically
generate the operator expansion order by order. We want to use this
insight as starting point for our study.

In Section~\ref{sec2} we introduce the necessary formal ingredients. The
 lowest  order in the
multiple scattering series is worked out in Section~\ref{sec3}. The next
term, which is second order in the NN t-operator,
is  constructed in Section~\ref{sec4}. In this order, the free 3N
propagator appears for the first time, leading to the notorious moving
logarithmic singularities. This can be avoided totally as shown 
 in~\cite{new_method1,new_method2}.
 We apply this new method in which the logarithmic singularities 
are replaced by single
poles, which then can be handled in a  similar fashion as
the poles in the 2N Lippmann-Schwinger equation. 
Having the second order
under control it is obvious to go to the next order. However,
we  will not  give this obvious continuation explicitly.
The inclusion of 3NF's, which
is most interesting from a physical point of view will be skipped in
this first formal attempt. However, we do not
expect  principal difficulty according to the experience for the 3N
bound state, where the inclusion of
3NF's has been worked out explicitly~\cite{2N3N}. The calculation of observables
based on the 3-dimensional form 
 is described in Section~\ref{sec5}. Various Appendices provide further
 information. Finally we conclude in
Section~\ref{sec6}.

\section{The formal ingredients}
\label{sec2}

Our standard form of the Faddeev equation for 3N scattering
is given by~\cite{report,actapolonica} 
\begin{eqnarray}
T \vert \phi  \rangle &=& tP\vert \phi  \rangle 
+ (1 + tG_0)V_4^{(1)}(1 + P)\vert \phi  \rangle
+ tPG_0T\vert \phi  \rangle  \cr
&+&(1 + tG_0)V_4^{(1)}(1+P)T\vert \phi  \rangle ~, 
\label{e1}
\end{eqnarray}
where $t$ is the t-operator for the NN pair 23, $G_0$ the free 
3N propagator, $P$ the sum of a cyclical and
an anticyclical permutation and $ V_4^ {(1)}$ the part of the 3NF
which is symmetrical under exchange of
nucleons 2 and 3. Here we arbitrarily choose nucleon 1 as being the spectator.

Knowing $ T \vert \phi  \rangle$ amplitudes for Nd elastic and 
breakup scattering 
are given by the matrix elements~\cite{report,book}
\begin{eqnarray}
\langle \Phi'| U | \Phi\rangle& =&  \langle \Phi'| P G_0^ {-1} 
+ PT | \Phi\rangle ~,
\label{e2}
\end{eqnarray}
\begin{eqnarray}
\langle \Phi_0| U_0| \Phi\rangle &=& \langle \Phi_0| ( 1 + P) T | \Phi\rangle ~.
\label{e3}
\end{eqnarray}
The state describing three free nucleons is given  by $| \Phi_0 \rangle$.

The iteration of Eq.~(\ref{e1}) generates the Faddeev multiple 
scattering series. When
neglecting 
3NFs, the iteration leads to
\begin{eqnarray}
T | \Phi\rangle = t P| \Phi\rangle + t P G_0  t P |\Phi \rangle + \cdots
\label{e4}
\end{eqnarray}
As we learned from 3-boson scattering driven by a 2-body force of 
 Malfliet-Tjion type~\cite{Liu:2004tv} 
which  incorporates  typical properties of the NN force, namely an
intermediate range attraction and a
short range repulsion, the convergence of the series from Eq.~(\ref{e4})
improves with increasing energy~\cite{Elster:2008yt}.

First, we introduce the three possible 3N isospin states
 $ | \gamma_a \rangle = | ( t_a \frac{1}{2} ) T_a M_T\rangle $:
\begin{eqnarray}
|\gamma_0 \rangle & = &  | ( 0 \frac{1}{2}) \frac{1}{2} M_T\rangle ~,\cr
|\gamma_1 \rangle & = &  | ( 1 \frac{1}{2}) \frac{1}{2} M_T\rangle ~,\cr
|\gamma_2 \rangle & = &  | ( 1 \frac{1}{2}) \frac{3}{2} M_T\rangle ~,
\label{e5-7}
\end{eqnarray}
in which the 2N isospin $ t $ is coupled with the isospin $\frac{1}{2}$ of 
the third particle to
the total isospin $ T= \frac{1}{2} $ or $ \frac{3}{2} $. As is well 
known~\cite{pd}  in  isospin space the 2N
t-operator has the form
\begin{eqnarray}
t = \sum_{ab} | \gamma_a\rangle  t_{ab} \langle \gamma_b| ~.
\label{e8}
\end{eqnarray}
We assume conservation of $ t_a $ but allow for charge independence
and 
charge symmetry breaking which
leads to the coupling of $ T = \frac{1}{2} $ and $ \frac{3}{2} $ states:
\begin{eqnarray}
t_{ab} = \delta_{ t_a t_b} t_{ t_a T_a T_b} ~.
\label{e9}
\end{eqnarray}
The linear combination of $ np $ t-operators in $ t =0 $ and $ t=1 $
states with pp (nn) t-operators
for proton-deuteron (pd) (neutron-deuteron (nd)) scattering is given 
in~\cite{isospin_t}. 
Furthermore, the permutation operator $P$ in the 3N isospin space 
reads~\cite{pd}
\begin{eqnarray}
\langle \gamma_a| P | \gamma_b\rangle = \delta_{ T_a T_b} F_{ t_a T_a T_b} 
( P_{12}^ {sm} P_{23}^ {sm} + ( - )^ { t_a + t_b} P_{13}^ {sm}
P_{23}^ {sm}) ~,
\label{e10}
\end{eqnarray}
where $ F_{ t_a T_a T_b } $ is essentially a $6j$-symbol~\cite{pd}
 and $ P_{ij}^ {sm} $ are transpositions of the 
nucleons $ij$ acting only in spin and momentum spaces. The  3N
momentum space is spanned by the
standard Jacobi momenta $ \vec p$ and $\vec q $~\cite{book}. Combining
spin/isospin space with the momentum space leads to the permutation operator
\begin{eqnarray}
\lefteqn{\langle \vec p \vec q | P_{12}^ {sm} P_{23}^ {sm} + ( -)^ { t_a 
+ t_b} P_{13}^ {sm} P_{23}^ {sm})
 | \vec p~' \vec q~'\rangle =}\cr
&    &    \delta( \vec p - \vec \pi( \vec q \vec q~'))
\delta( \vec p~' - \vec \pi'( \vec q \vec q~')) P_{12}^s P_{23}^ s 
 + ( -)^ { t_a + t_b} \delta( \vec p + \vec \pi( \vec q \vec q~'))
\delta( \vec p~' + \vec \pi'( \vec q \vec q~'))P_{13}^ s P_{23}^ s
\label{e11}
\end{eqnarray}
with
\begin{eqnarray}
\vec \pi( \vec q, \vec q~') & = &  \frac{1}{2} \vec q + \vec q~' ~, \cr
 \vec \pi'( \vec q, \vec q~') & = &  -  \vec q - \frac{1}{2}  \vec
  q~' ~.
\label{e13}
\end{eqnarray}

Next we use the operator form expansion of the off-shell NN 
t-operator~\cite{2N3N,new}
 in the 3N spin and momentum spaces
\begin{eqnarray}
\langle \vec p \vec q| t_{ t_a T_a T_b} | \vec p~' \vec q~'\rangle  
=   \sum_j t_{ t_a T_a T_b}^ { (j)}( \vec p,
\vec p~', E_q)  w_j ( \vec \sigma(2), \vec \sigma(3), \vec p,  \vec p~') 
\delta( \vec q - \vec q~') ~,
\label{e14}
\end{eqnarray}
where $ E_q = E - \frac{3}{4m} q^ 2 $ and $E$ the total c.m. energy.
Due to parity and time reversal invariance exactly 6 terms of scalar 
 spin-momentum operators $w_j$ are
possible~\cite{wolfenstein}. They are
\begin{eqnarray}
w_1 ( \vec \sigma(1),\vec \sigma(2), \vec p, \vec p~')&  = &  1\cr
w_2 ( \vec \sigma(1),\vec \sigma(2), \vec p, \vec p~')&  = & 
\vec \sigma(1) \cdot \vec \sigma(2)\cr
w_3 ( \vec \sigma(1),\vec \sigma(2), \vec p, \vec p~')&  = &  
( \vec \sigma(1) + \vec \sigma(2) ) \cdot (
\vec p \times \vec p~')\cr
w_4 ( \vec \sigma(1),\vec \sigma(2), \vec p, \vec p~')&  = & 
\vec \sigma(1) \cdot ( \vec p \times \vec p~')
\vec \sigma(2) \cdot ( \vec p \times \vec p~')\cr
w_5 ( \vec \sigma(1),\vec \sigma(2), \vec p, \vec p~')&  = & 
\vec \sigma(1) \cdot ( \vec p + \vec p~') \vec
\sigma (2) \cdot ( \vec p + \vec p~')\cr
w_6 ( \vec \sigma(1),\vec \sigma(2), \vec p, \vec p~')&  = & 
\vec \sigma(1) \cdot ( \vec p - \vec p~') \vec
\sigma(2) \cdot ( \vec p - \vec p~').
\label{e15-20}
\end{eqnarray}
Using this operator representation of the NN force, the functions  
$ t_{ t_a T_a T_b}^ { (j)}( \vec p, \vec p~', E_q)$ of Eq.~(\ref{e14}) are scalar
functions and depend only on
 three variables, the magnitudes of the vectors $\vec p$ and $\vec
 p~'$, 
and the angle
between them, $\hat p \cdot \hat p~'$.

Finally, we use the operator form of the initial state as given 
in Ref.~\cite{fachruddin}
\begin{eqnarray}
\lefteqn{\langle \vec p \vec q| \langle \gamma_0| \Phi\rangle 
\equiv \langle \vec p \vec q| \phi\rangle }\cr
 &=&   ( \phi_1( p) + \phi_2( p) ( \vec \sigma(2) \cdot \vec p \vec
 \sigma(3) \cdot \vec p   - \frac{1}{3} p^ 2) | 1 m_d\rangle | m_{10}\rangle 
\delta( \vec q - \vec q_0) \cr
&  \equiv & \sum_{k=1}^ 2 \phi_k( p) O_k( \vec \sigma(2), 
\vec \sigma(3), \vec p) | 1 m_d\rangle | m_{10}\rangle
 \delta( \vec q - \vec q_0) ~.
\label{e22}
\end{eqnarray}
Here $ \phi_1(p)$ and $\phi_2(p)$ are proportional to the standard s-
and d-wave components
 of the deuteron wave function. The state  $| 1 m_d\rangle $ is the
 pure 
spin 1 two-nucleon
 state 
with spin magnetic quantum 
number $ m_d$, and $ m_{10}$ is the initial state spin magnetic quantum
number of 
the projectile nucleon.
 The vector  $ \vec q_0 $ is the initial relative  momentum in the Nd system.
All these are the necessary ingredients to work out the terms in the 
multiple scattering series of Eq.~(\ref{e4}).

\section{The first order in $t$}
\label{sec3}

We consider the first order term in Eq.~(\ref{e4}) and use Eqs.~(\ref{e8}) -
(\ref{e11}), (\ref{e14}) and (\ref{e22}) to obtain
\begin{eqnarray}
 \langle \vec p \vec q |\langle \gamma_a|  t P | \Phi\rangle
  =   F_{ t_a t_0 T_0}
 \langle \vec p \vec q| t_{t_a T_a T_0}  ( P_{12}^ {sm} P_{23}^ {sm} + ( -)^ {
t_a + t_0} P_{13}^ {sm} P_{23}^ {sm})| \phi\rangle ~.
\label{e23}
\end{eqnarray}
Here we used the isospin property of the initial state:  $ t_0 =0, T_0 = 1/2$.
Then
\begin{eqnarray}
\lefteqn{\langle \vec p \vec q |\langle \gamma_a|  t P | \Phi\rangle} \cr
&  = &  F_{ t_a t_0 T_0}\int d^ 3 p' d^ 3 q' \sum_j t_{ t_a T_a T_0}^
  { (j)}
( \vec p, \vec p~',E_q)
 w_j ( \vec \sigma(2), \vec \sigma(3), \vec p,  \vec p~')  \delta( \vec q 
- \vec q~')\cr
& &  \int d^ 3 p'' d^ 3 q'' (  \delta( \vec p~' - \vec \pi( \vec q~' \vec q~''))
\delta( \vec p~'' - \vec \pi'( \vec q~' \vec q~'')) P_{12}^s P_{23}^ s\cr
&  + &  ( -)^ { t_a + t_0} \delta( \vec p~' + \vec \pi( \vec q~' \vec q~''))
\delta( \vec p~'' + \vec \pi'( \vec q~' \vec q~''))P_{13}^ s P_{23}^ s)\cr
& & \sum_{k=1}^ 2 \phi_k( p'') O_k( \vec \sigma(2), \vec \sigma(3), \vec
p~'') 
| 1 m_d\rangle | m_{10}\rangle \delta(
\vec q~'' - \vec q_0)\cr
& = & F_{ t_a t_0 T_0}[ \sum_j t_{ t_a T_a T_0}^ { (j)}( \vec p,  \vec
  \pi
( \vec q \vec q_0),E_q )
w_j ( \vec \sigma(2), \vec \sigma(3), \vec p,  \vec \pi( \vec q \vec
q_0)) 
P_{12}^ s P_{23}^ s\cr
& & \sum_{k=1}^ 2 \phi_k( | \vec \pi'( \vec q \vec q_0)|) O_k( \vec
\sigma(2), 
\vec \sigma(3), \vec \pi'(
\vec q \vec q_0)) | 1 m_d\rangle | m_{10}\rangle\cr
& + & ( -)^ { t_a + t_0} \sum_j t_{ t_a T_a T_0}^ { (j)}( \vec p ,-
\vec  \pi( 
\vec q \vec q_0),E_q)
w_j ( \vec \sigma(2), \vec \sigma(3), \vec p , - \vec \pi( \vec q 
\vec q_0))P_{13}^ s P_{23}^ s\cr
& & \sum_{k=1}^ 2 \phi_k(| \vec \pi'( \vec q \vec q_0)| ) O_k( 
\vec \sigma(2), \vec \sigma(3), \vec \pi'(
\vec q \vec q_0)) | 1 m_d\rangle | m_{10}\rangle] ~.
\label{e24}
\end{eqnarray}
Note that we used the fact that the two operators $O_k$ from Eq.~(\ref{e22})
 depend quadratically on the momenta. 
Then we  define
\begin{eqnarray}
\lefteqn{ w_j ( \vec \sigma(2), \vec \sigma(3), \vec p, \vec \pi) 
P_{12}^ s P_{23}^ s 
O_k( \vec \sigma(2), \vec \sigma(3),  \vec \pi')} \cr
&  = &  w_j ( \vec \sigma(2), \vec \sigma(3), \vec p ,\vec \pi)
O_k( \vec \sigma(3), \vec \sigma(1),  \vec \pi') P_{12}^ s P_{23}^ s\cr
& \equiv& a_{jk} ( \vec \sigma(1), \vec \sigma(2), \vec \sigma(3),  
\vec p, \vec \pi, \vec \pi')
P_{12}^ s P_{23}^ s ~, 
\label{e25}
\end{eqnarray}
and
\begin{eqnarray}
\lefteqn{  w_j ( \vec \sigma(2),  \vec \sigma(3),  \vec p, 
-  \vec \pi) P_{13}^ s P_{23}^ s
O_k( \vec \sigma(2), \vec \sigma(3),  \vec \pi') }\cr
&  = &  w_j ( \vec \sigma(2), \vec \sigma(3), \vec p,- \vec \pi)
O_k( \vec \sigma(1), \vec \sigma(2),   \vec \pi')P_{13}^ s P_{23}^ s \cr
& \equiv&  b_{jk}( \vec \sigma(1), \vec \sigma(2), \vec \sigma(3),  
\vec p,  \vec \pi, \vec \pi')
P_{13}^ s P_{23}^ s ~.
\label{e26} 
\end{eqnarray}
The scalar expressions $ a_{ jk} $ and $ b_{jk} $  have to be worked 
out such  that  each $ \vec \sigma(i)$
 occurs only once. The explicit expressions are given in  Appendix~\ref{ap1}.

Inserting then Eqs.~(\ref{e25}) and (\ref{e26}) into Eq.~(\ref{e24}) yields
\begin{eqnarray}
\lefteqn{\langle \vec p \vec q |\langle \gamma_a|  t P | \Phi\rangle} \cr
&  = & \sum_{jk} t_{ t_a T_a T_0}^ { (j)}( \vec p,  \vec  \pi, E_q )
  \phi_k
( | \vec \pi'|)F_{ t_a t_0 T_0}\cr
& &  a_{jk} ( \vec \sigma(1), \vec \sigma(2), \vec \sigma(3), \vec p, 
\vec \pi, \vec
\pi')P_{12}^ s P_{23}^ s  | 1 m_d\rangle | m_{10}\cr
& + & \sum_{jk} t_{ t_a T_a T_0}^ { (j)}( \vec p, -  \vec  \pi, E_q ) 
\phi_k( | \vec \pi'|)F_{ t_a t_0
T_0} ( -)^ { t_a + t_0}\cr
& &  b_{jk} ( \vec \sigma(1), \vec \sigma(2), \vec \sigma(3), \vec p, 
\vec \pi, \vec \pi')
P_{13}^ s P_{23}^ s | 1 m_d\rangle | m_{10} \rangle ~.
\label{e27}
\end{eqnarray}
We see the expected structure, a sum over the product of scalar
operators   
$( a_{jk}, b_{jk})$ multiplied by 
scalar functions. 
Note that
\begin{eqnarray}
a_{jk} & = &  a_{jk}( \vec \sigma(1), \vec \sigma(2), \vec \sigma(3), 
\vec p, \vec \pi( \vec q \vec q_0), 
\vec \pi' ( \vec q \vec q_0)) ~,\cr
b_{jk} & = &  b_{jk}( \vec \sigma(1), \vec \sigma(2), \vec \sigma(3), 
\vec p, \vec \pi( \vec q \vec q_0),
\vec \pi' ( \vec q \vec q_0)) ~.
\label{e28-29}
\end{eqnarray}
It remains to display the singularity structure for the calculation 
of the physical amplitudes and the
treatment of the second order.

The NN t-matrix in the pd isospin space has the general structure
\begin{eqnarray}
t_{ t_a T_a T_b}  = \delta_{ t_a 0} t_{ 0 \frac{1}{2} \frac{1}{2}} 
+ \delta_{ t_a 1} t_{ 1 T_a T_b} ~,
\label{e30}
\end{eqnarray}
where
\begin{eqnarray}
t_{ 0 \frac{1}{2} \frac{1}{2}} & = &  t_{np}^ {00} ~, \\
t_{ 1 T_a T_b} & = &   \alpha_{ T_a T_b } t_{np}^ {10} 
+ \beta_{ T_a T_b } t_{pp}^ {1 -1} ~,
\label{e31-32}
\end{eqnarray}
and $ \alpha_{ T_a T_b }, \beta_{ T_a T_b }$ are numerical 
values related to Clebsch Gordon coefficients.

\noindent
The np t-matrix $ t_{np}^ {00} $ for $ t=0 $ has a deuteron pole
\begin{eqnarray}
t_{np}^ {00}( \vec p, \vec p~', E_q) \equiv \frac{ \hat t_{np}^ {00}( 
\vec p, \vec p~', E_q) }{ E_q + i
\epsilon  - E_d} ~.
\label{e33}
\end{eqnarray}
Therefore, Eq.~(\ref{e27}) can be decomposed further,
\begin{eqnarray}
\lefteqn{\langle \vec p \vec q |\langle \gamma_a|  t P | \Phi\rangle } \cr
&  = & \sum_{jk}
( \delta_{ t_a 0} \frac{ \hat t_{np}^ {00, (j)}( \vec p,  \vec  \pi,
      E_q )}
{ E_q + i \epsilon  - E_d}
+ \delta_{ t_a 1} t_{ 1 T_a T_0}^ { (j)}( \vec p,  \vec  \pi, E_q ))
 \phi_k( | \vec \pi'|)F_{ t_a t_0 T_0}\cr
& &  a_{jk} ( \vec \sigma(1), \vec \sigma(2), \vec \sigma(3), 
\vec p, \vec \pi, \vec
\pi') P_{12}^ s P_{23}^ s | 1 md\rangle | m_{10} \rangle\cr
& + & \sum_{jk}
( \delta_{ t_a 0} \frac{ \hat t_{np}^ {00, (j)}( \vec p, 
- \vec  \pi, E_q )}{ E_q + i \epsilon  - E_d}
+ \delta_{ t_a 1} t_{ 1 T_a T_0}^ { (j)}( \vec p,  - \vec  \pi, E_q ))
 \phi_k( | \vec \pi'|)F_{ t_a t_0 T_0}( -)^ { t_a + t_0}\cr
& &  b_{jk} ( \vec \sigma(1), \vec \sigma(2), \vec \sigma(3), 
\vec p, \vec \pi, \vec \pi')
 P_{13}^ s P_{23}^ s | 1 md\rangle | m_{10} \rangle ~,
\label{e34}
\end{eqnarray}
which retains the structure of scalar spin-momentum dependent operators 
and momentum dependent scalar functions.

\section{ The second order in $t$}
\label{sec4}

Next we consider the second order term in Eq.~(\ref{e4}) and perform 
the isospin 
projections according to Refs.~(\ref{e8}) and (\ref{e10})
\begin{eqnarray}
\langle \gamma_a| t P G_0 t P | \Phi\rangle & =  & t_{ab} \langle \gamma_b| P | 
\gamma_c\rangle G_0 \langle \gamma_c | tP | \Phi\rangle\cr
&  = &  \delta_{ t_a t_b} t_{ t_a T_a T_b} \delta_{ T_b T_c} F_{ t_b t_c T_b}
 ( P_{12}^ {sm} P_{23}^ {sm} + ( -)^ { t_b + t_c} P_{13}^ {sm} P_{23}^
      {sm}) G_0 \langle \gamma_c | tP | \Phi\rangle\cr
&  = &  F_{ t_a t_c T_b} t_{ t_a T_a T_b}( P_{12}^ {sm} P_{23}^ {sm}
+ ( -)^ { t_a + t_c} P_{13}^ {sm} P_{23}^ {sm}) G_0 \delta_{ T_b T_c}
\langle \gamma_c | tP | \Phi\rangle ~.
\label{e35}
\end{eqnarray}
Then we insert Eqs.~(\ref{e14}) and (\ref{e11}) 
\begin{eqnarray}
\lefteqn{\langle \vec p \vec q| \langle \gamma_a| t P G_0 t P | \Phi\rangle }\cr
&   = &  F_{ t_a t_c T_b} \int d^ 3 p' \sum_j t^ {(j)}_{ t_a T_a T_b} 
( \vec p, \vec p~', E_q)
 w_j( \vec \sigma(2), \vec \sigma(3), \vec p, \vec p~')\cr
& &  \int d^ 3 p_2 d^ 3 q_2 \langle \vec p~' \vec q | ( P_{12}^ {sm} P_{23}^
 {sm} 
+ ( -)^ { t_a + t_c} P_{13}^
{sm} P_{23}^ {sm})| \vec p_2 \vec q_2\rangle G_0( \vec p_2 \vec q_2)\cr
& &  \delta_{ T_b T_c} \langle \vec p_2 \vec q_2 
| \langle \gamma_c | tP | \Phi\rangle\cr
& = & F_{ t_a t_c T_b} \int d^ 3 p' \sum_j t^ {(j)}_{ t_a T_a T_b} (
\vec p, 
\vec p~', E_q)
 w_j( \vec \sigma(2), \vec \sigma(3), \vec p, \vec p~')\cr
& &  \int d^ 3 p_2 d^ 3 q_2 ( \delta( \vec p~' - \vec \pi( \vec q \vec q_2))
\delta( \vec p_2 - \vec \pi' ( \vec q \vec q_2)) P_{12}^ s P_{23}^ s\cr
&  + &  ( -)^ { t_a + t_c} \delta( \vec p~' + \vec \pi( \vec q \vec q_2))
\delta( \vec p_2 + \vec \pi' ( \vec q \vec q_2)) P_{13}^ s P_{23}^ s)\cr
& &  G_0( \vec p_2 \vec q_2)  \delta_{ T_b T_c} \langle \vec p_2 \vec q_2 | 
\langle \gamma_c | tP | \Phi\rangle\cr
& = & F_{ t_a t_c T_b} \sum_j  \int  d^ 3 q_2 \; \Big[ 
 t^ {(j)}_{ t_a T_a T_b} ( \vec p, \vec \pi( \vec q \vec q_2), E_q)
w_j( \vec \sigma(2), \vec \sigma(3), \vec p, \vec \pi( \vec q \vec q_2))\cr
& &  P_{12}^ s P_{23}^ s G_0( \vec \pi' ( \vec q \vec q_2) \vec q_2)
 \delta_{T_b T_c} \langle \vec \pi' ( \vec q \vec q_2) \vec q_2 | 
\langle \gamma_c | tP | \Phi\rangle\cr
&  + &  ( -)^ { t_a + t_c} t^ {(j)}_{ t_a T_a T_b} ( \vec p, - \vec \pi
( \vec q \vec q_2), E_q)
w_j( \vec \sigma(2), \vec \sigma(3), \vec p, -\vec \pi( \vec q \vec q_2))\cr
& & P_{13}^ s P_{23}^ s  G_0(\vec \pi' ( \vec q \vec q_2) \vec q_2)
\delta_{ T_b T_c} \langle - \vec \pi' ( \vec q \vec q_2) \vec q_2 | 
\langle \gamma_c | tP | \Phi\rangle \Big] ~.
\label{e36}
\end{eqnarray}
Finally we use the expression from Eq.~(\ref{e27}) for the first order term
\begin{eqnarray}
\lefteqn{ \langle \vec p \vec q| \langle \gamma_a| t P G_0 t P 
| \Phi\rangle }\cr
&   = & F_{ t_a t_c T_b} \sum_j  \int  d^ 3 q_2 \; \Big[ 
 t^ {(j)}_{ t_a T_a T_b} ( \vec p, \vec \pi( \vec q \vec q_2), E_q)
w_j( \vec \sigma(2), \vec \sigma(3), \vec p, \vec \pi( \vec q \vec q_2))\cr
& &  P_{12}^ s P_{23}^ s G_0( \vec \pi' ( \vec q \vec q_2) \vec q_2) 
\delta_{T_b T_c}\cr
& & \sum_{lk}( t_{ t_c T_b T_0}^ { (l)}(\vec \pi' ( \vec q \vec q_2),
\vec \pi ( \vec q_2,\vec q_0),
E_{q_2})
\phi_k( | \vec \pi'( \vec q_2,\vec q_0)|)F_{ t_c t_0 T_0}\cr
& & a_{lk} ( \vec \sigma(1), \vec \sigma(2), \vec \sigma(3), 
\vec \pi' ( \vec q \vec q_2),
\vec \pi( \vec q_2,\vec q_0), \vec \pi'( \vec q_2,\vec q_0))
P_{12}^ s P_{23}^ s | 1 m_d\rangle | m_{10} \rangle\cr
& + &  t_{ t_c T_b T_0}^ { (l)}( \vec \pi' ( \vec q \vec q_2), 
-  \vec  \pi (\vec q_2,\vec q_0),
E_{q_2} ) \phi_k ( | \vec \pi'( \vec q_2,\vec q_0)|) F_{ t_c t_0 T_0} 
(-)^{ t_c + t_0}\cr
& & b_{lk} ( \vec \sigma(1), \vec \sigma(2), \vec \sigma(3), \vec \pi' 
( \vec q \vec q_2), \vec \pi( \vec
q_2 \vec q_0) , \vec \pi' ( \vec q_2 \vec q_0))P_{13}^ s P_{23}^ s 
| 1 m_d\rangle | m_{10} \rangle)\cr
&  + &  ( -)^ { t_a + t_c} t^ {(j)}_{ t_a T_a T_b} ( \vec p, - \vec
\pi( 
\vec q \vec q_2), E_q)
w_j( \vec \sigma(2), \vec \sigma(3), \vec p, -\vec \pi( \vec q \vec q_2))\cr
& & P_{13}^ s P_{23}^ s  G_0(\vec \pi' ( \vec q \vec q_2) \vec q_2) 
\delta_{ T_b T_c}\cr
& & \sum_{lk}( t_{ t_c T_b T_0}^ { (l)}( - \vec \pi' ( \vec q \vec
q_2),
\vec  \pi ( \vec q_2,\vec q_0),
E_{q_2})
 \phi_k( | \vec \pi'( \vec q_2,\vec q_0)|)F_{ t_c t_0 T_0}\cr
& & a_{lk} ( \vec \sigma(1), \vec \sigma(2), \vec \sigma(3),
- \vec \pi' ( \vec q \vec q_2),
\vec \pi( \vec q_2,\vec q_0), \vec \pi'( \vec q_2,\vec q_0)) 
P_{12}^ s P_{23}^ s| 1 m_d\rangle | m_{10} \rangle\cr
& + &  t_{ t_c T_b T_0}^ { (l)}(- \vec \pi' ( \vec q \vec q_2), 
-  \vec  \pi (\vec q_2,\vec q_0),
E_{q_2})
\phi_k ( | \vec \pi'( \vec q_2,\vec q_0)|) F_{ t_c t_0 T_0} 
(-)^ { t_c + t_0}\cr
& & b_{lk} ( \vec \sigma(1), \vec \sigma(2), \vec \sigma(3), 
- \vec \pi' ( \vec q \vec q_2),
\vec \pi( \vec q_2 \vec q_0) , \vec \pi' ( \vec q_2 \vec q_0))\cr
& & P_{13}^ s P_{23}^ s | 1 m_d\rangle | m_{10} \rangle) \Big] ~.
\label{e37}
\end{eqnarray}

\noindent
Now we collect the spin parts and define
\begin{eqnarray}
\lefteqn{ C_{ j l k} (  \vec \sigma(1), \vec \sigma(2), \vec \sigma(3),
\vec p, \vec \pi( \vec q \vec q_2),
\vec  \pi'( \vec q \vec q_2),  \vec \pi( \vec q_2 \vec q_0), 
\vec \pi'( \vec q_2 \vec q_0) )  P_{12}^ s P_{23}^ s } \cr
& \equiv &  w_j( \vec \sigma(2), \vec \sigma(3), \vec p,\vec   \pi( 
\vec q \vec q_2))  P_{12}^ s P_{23}^ s
 a_{ lk} ( \vec \sigma(1), \vec \sigma(2), \vec \sigma(3), 
\vec  \pi'( \vec q \vec q_2),
\vec \pi( \vec q_2 \vec q_0), \vec \pi'( \vec q_2 \vec q_0))\cr
& = & w_j( \vec \sigma(2), \vec \sigma(3), \vec p,\vec  \pi
( \vec q \vec q_2))
a_{ l k} ( \vec \sigma(2), \vec \sigma(3), \vec \sigma(1),
\vec \pi' ( \vec q \vec q_2),  \vec \pi( \vec q_2 \vec q_0), 
\vec \pi'( \vec q_2 \vec q_0) )
  P_{12}^ s P_{23}^ s ~,
\label{e38}
\end{eqnarray}
\begin{eqnarray}
\lefteqn{D_{ j l k} (  \vec \sigma(1), \vec \sigma(2), \vec \sigma(3),
\vec p, \vec \pi( \vec q \vec q_2),
\vec \pi'( \vec q \vec q_2),  \vec \pi( \vec q_2 \vec q_0), 
\vec \pi'( \vec q_2 \vec q_0) )   P_{12}^ s P_{23}^ s } \cr
& \equiv &  w_j( \vec \sigma(2), \vec \sigma(3), \vec p, 
\vec  \pi( \vec q \vec q_2))  P_{12}^ s P_{23}^ s
b_{ l k} ( \vec \sigma(1), \vec \sigma(2), \vec \sigma(3), 
\vec  \pi'( \vec q \vec q_2),
  \vec \pi( \vec q_2 \vec q_0), \vec \pi'( \vec q_2 \vec q_0) )\cr
& = & w_j( \vec \sigma(2), \vec \sigma(3), \vec p, \vec  \pi
( \vec q \vec q_2))
b_{ l k} ( \vec \sigma(2), \vec \sigma(3), \vec \sigma(1), 
\vec  \pi'( \vec q \vec q_2),
 \vec \pi( \vec q_2 \vec q_0), \vec \pi'( \vec q_2 \vec q_0) )
 P_{12}^ s P_{23}^ s ~,
\label{e39}
\end{eqnarray}
\begin{eqnarray}
\lefteqn{ E_{ j l k} (  \vec \sigma(1), \vec \sigma(2), \vec \sigma(3), 
\vec p, \vec  \pi( \vec q \vec q_2),
\vec  \pi'( \vec q \vec q_2),  \vec \pi( \vec q_2 \vec q_0), 
\vec \pi'( \vec q_2 \vec q_0) ) P_{13}^ s P_{23}^ s } \cr
&  \equiv &  w_j( \vec \sigma(2), \vec \sigma(3), \vec p, - 
\vec  \pi( \vec q \vec q_2)) P_{13}^ s P_{23}^ s
a_{ lk} ( \vec \sigma(1), \vec \sigma(2), \vec \sigma(3), 
- \vec \pi'( \vec q \vec q_2),
\vec \pi( \vec q_2 \vec q_0), \vec \pi'( \vec q_2 \vec q_0) )\cr
& = & w_j( \vec \sigma(2), \vec \sigma(3), \vec p, - \vec  \pi
( \vec q \vec q_2))
a_{ lk} ( \vec \sigma(3), \vec \sigma(1), \vec \sigma(2),
- \vec \pi'( \vec q \vec q_2),
\vec \pi( \vec q_2 \vec q_0), \vec \pi'( \vec q_2 \vec q_0) )
 P_{13}^ s P_{23}^ s ~,
\label{e40}
\end{eqnarray}
\begin{eqnarray}
\lefteqn{ F_{ j l k} (  \vec \sigma(1), \vec \sigma(2), 
\vec \sigma(3), \vec p, \vec  \pi( \vec q \vec q_2),
\vec \pi'( \vec q \vec q_2),  \vec \pi( \vec q_2 \vec q_0), 
\vec \pi'( \vec q_2 \vec q_0) ) P_{13}^ s P_{23}^ s  } \cr
& \equiv & w_j( \vec \sigma(2), \vec \sigma(3), \vec p,
- \vec \pi( \vec q \vec q_2))  P_{13}^ s P_{23}^ s
b_{ l k} ( \vec \sigma(1), \vec \sigma(2), \vec \sigma(3), 
- \vec \pi'( \vec q \vec q_2),
\vec \pi( \vec q_2 \vec q_0), \vec \pi'( \vec q_2 \vec q_0) )\cr
& = & w_j( \vec \sigma(2), \vec \sigma(3), \vec p, - \vec \pi( \vec q \vec q_2))
b_{ l k} ( \vec \sigma(3), \vec \sigma(1), \vec \sigma(2),
- \vec \pi'( \vec q \vec q_2),
\vec \pi( \vec q_2 \vec q_0), \vec \pi'( \vec q_2 \vec q_0) )
 P_{13}^ s P_{23}^ s ~.
\label{e41}
\end{eqnarray}
Inserting these expressions into Eq.~(\ref{e37}) yields
\begin{eqnarray}
\lefteqn{\langle \vec p \vec q| \langle \gamma_a| t P G_0 t P 
| \Phi\rangle } \cr
& = & F_{ t_a t_c T_b} \sum_j  \int  d^ 3 q_2 \; \Big[ 
 t^ {(j)}_{ t_a T_a T_b} ( \vec p, \vec \pi( \vec q \vec q_2), E_q)
G_0( \vec \pi' ( \vec q \vec q_2) \vec q_2) \delta_{T_b T_c}\cr
& & \sum_{lk}( t_{ t_c T_b T_0}^ { (l)}(\vec \pi' ( \vec q \vec q_2),
\vec \pi ( \vec q_2,\vec q_0),
E_{q_2})\phi_k( | \vec \pi'( \vec q_2,\vec q_0)|)F_{ t_c t_0 T_0}\cr
& & C_{ j l k} (  \vec \sigma(1), \vec \sigma(2), \vec \sigma(3),
\vec p, \vec \pi( \vec q \vec q_2),
\vec  \pi'( \vec q \vec q_2),  \vec \pi( \vec q_2 \vec q_0), 
\vec \pi'( \vec q_2 \vec q_0) ) P_{13}^ s P_{23}^ s \cr
& + &  t_{ t_c T_b T_0}^ { (l)}( \vec \pi' ( \vec q \vec q_2), 
-  \vec  \pi (\vec q_2,\vec q_0),
E_{q_2} ) \phi_k ( | \vec \pi'( \vec q_2,\vec q_0)|) F_{ t_c t_0 T_0} 
(-)^ { t_c + t_0}\cr
& & D_{ j l k} (  \vec \sigma(1), \vec \sigma(2), \vec \sigma(3),\vec
p, 
\vec \pi( \vec q \vec q_2),
\vec  \pi'( \vec q \vec q_2),  \vec \pi( \vec q_2 \vec q_0), \vec \pi'
( \vec q_2 \vec q_0) )  )\cr
&  + &  ( -)^ { t_a + t_c} t^ {(j)}_{ t_a T_a T_b} ( \vec p, - \vec
\pi
( \vec q \vec q_2), E_q)
G_0(\vec \pi' ( \vec q \vec q_2) \vec q_2) \delta_{ T_b T_c}\cr
& & \sum_{lk}( t_{ t_c T_b T_0}^ { (l)}( - \vec \pi' ( \vec q \vec
q_2),
\vec  \pi ( \vec q_2,\vec q_0),
E_{q_2})
 \phi_k( | \vec \pi'( \vec q_2,\vec q_0)|)F_{ t_c t_0 T_0}\cr
& & E_{ j l k} (  \vec \sigma(1), \vec \sigma(2), \vec \sigma(3),
\vec p, \vec \pi( \vec q \vec q_2),
\vec  \pi'( \vec q \vec q_2),  \vec \pi( \vec q_2 \vec q_0), \vec \pi'
( \vec q_2 \vec q_0) ) \cr
& + &  t_{ t_c T_b T_0}^ { (l)}(- \vec \pi' ( \vec q \vec q_2), -  
\vec  \pi (\vec q_2,\vec q_0),
E_{q_2})
\phi_k ( | \vec \pi'( \vec q_2,\vec q_0)|) F_{ t_c t_0 T_0} ( -)^ { t_c + t_0}\cr
& & E_{ j l k} (  \vec \sigma(1), \vec \sigma(2), \vec \sigma(3),\vec
p, 
\vec \pi( \vec q \vec q_2),
\vec  \pi'( \vec q \vec q_2),  \vec \pi( \vec q_2 \vec q_0), \vec \pi'
( \vec q_2 \vec q_0) )\cr
& &  P_{12}^ s P_{23}^ s)\Big] | 1 m_d\rangle | m_{10} \rangle ~.
\label{e42}
\end{eqnarray}

As will be shown for  some examples in Appendix B  the coefficients 
$C_{ j l k}$ from Eq.~(\ref{e38}) together with the other coefficients 
from Eqs.~(\ref{e39}) through (\ref{e41})
 will depend on the following types of scalars. As an aside, one more term,
$ \vec \sigma(i_1) \cdot \vec q_2
( \vec \sigma(i_2) \times \vec \sigma(i_3) ) \cdot \vec q_2 $, 
is expected to occur in $ C_{ jlk} $ beyond
the ones given in Appendix C:
\begin{eqnarray}
\lefteqn{ C_{ j l k} (  \vec \sigma(1), \vec \sigma(2), \vec \sigma(3),\vec p, 
\vec \pi( \vec q \vec q_2),
\vec  \pi'( \vec q \vec q_2),  \vec \pi( \vec q_2 \vec q_0), \vec \pi'
( \vec q_2 \vec q_0) ) }\cr
& = &  C_{ j l k}^ {(a)} ( \vec \sigma(1), \vec \sigma(2), \vec
  \sigma(3),
\vec p, \vec q, \vec q_0)
  +   C_{ j l k}^ {(b)}( \vec p, \vec q, \vec q_0, \vec q_2)\cr
&  + &  \sum_{ i_1} C_{ j l k; i_1}^ {(c)} ( \vec \sigma( i_2), 
\vec \sigma( i_3), \vec p, \vec q,\vec q_0)
 \vec \sigma( i_1) \cdot \vec q_2\cr
&  + & \sum_{ i_1 i_2} C_{ j l k; i_1 i_2}^ {(d)} ( \vec \sigma( i_3), 
\vec p,\vec q,\vec q_0) \vec
\sigma( i_1) \cdot \vec q_2 \vec \sigma( i_2) \cdot \vec q_2\cr
& + & C_{ j l k}^ {(e)}( \vec p,\vec q,\vec q_0) \vec \sigma( i_1) 
\cdot \vec q_2 \vec \sigma( i_2) \cdot 
\vec q_2  \sigma( i_3) \cdot \vec q_2 \cr
& + & \sum_{ i_1 \ne i_2} C_{ j l k; i_1 i_2}^ {(f)} ( \vec \sigma(
i_3) , 
\vec p, \vec q,\vec q_0) \vec
\sigma( i_1 ) \times \vec \sigma( i_2) \cdot \vec q_2\cr
& + & \sum_{ i_1 \ne i_2 \ne i_3} C_{ jlk; i_1 i_2 i_3}^ {(g)} ( \vec p, \vec
q, \vec q_0), \vec \sigma( i_1 ) \cdot \vec q_2 (  \vec \sigma( i_2 ) 
\times \vec \sigma( i_3)) \cdot \vec q_2) ~, 
\label{e43}
\end{eqnarray}
where $ i_1, i_2, i_3 = 1,2,3 $ and cyclical permutations thereof.

It is sufficient to consider the first integral in Eq.~(\ref{e37}) 
going along with the coefficients $ C_{jlk}$. 
We insert Eq.~(\ref{e43}):
\begin{eqnarray}
\lefteqn{\langle \vec p \vec q| \langle \gamma_a| t P G_0 t P 
| \Phi\rangle^ {(1)} }\cr
& = & F_{ t_a t_c T_b}\delta_{T_b T_c}F_{ t_c t_0 T_0} \sum_j\sum_{lk} \;
 \Big[ C_{ j l k}^ {(a)} ( \vec \sigma(1), \vec \sigma(2), \vec \sigma(3),
\vec p, \vec q, \vec q_0)\cr
& &  \int  d^ 3 q_2 t^ {(j)}_{ t_a T_a T_b} ( \vec p, \vec \pi( \vec q 
\vec q_2), E_q)
G_0( \vec \pi' ( \vec q \vec q_2) \vec q_2)\cr
& & t_{ t_c T_b T_0}^ { (l)}(\vec \pi' ( \vec q \vec q_2),\vec \pi 
( \vec q_2,\vec q_0),
E_{q_2})\phi_k( | \vec \pi'( \vec q_2,\vec q_0)|) \cr
& + & \int  d^ 3 q_2 t^ {(j)}_{ t_a T_a T_b} ( \vec p, \vec \pi( 
\vec q \vec q_2), E_q)
 G_0( \vec \pi' ( \vec q \vec q_2) \vec q_2)\cr
& & t_{ t_c T_b T_0}^ { (l)}(\vec \pi' ( \vec q \vec q_2),\vec \pi 
( \vec q_2,\vec q_0),
E_{q_2})\phi_k( | \vec \pi'( \vec q_2,\vec q_0)|) C_{ j l k}^ {(b)}
( \vec p, \vec q, \vec q_0, \vec q_2)\cr
& + & \sum_{ i_1} C_{ j l k; i_1}^ {(c)} ( \vec \sigma( i_2), 
\vec \sigma( i_3), \vec p,\vec q,\vec q_0)
 \int  d^ 3 q_2 t^ {(j)}_{ t_a T_a T_b} ( \vec p, \vec \pi
( \vec q \vec q_2), E_q)
G_0( \vec \pi' ( \vec q \vec q_2) \vec q_2)\cr
& & t_{ t_c T_b T_0}^ { (l)}(\vec \pi' ( \vec q \vec q_2),
\vec \pi ( \vec q_2,\vec q_0),
E_{q_2})\phi_k( | \vec \pi'( \vec q_2,\vec q_0)|) \vec \sigma( i_1) 
\cdot \vec q_2\cr
& + & \sum_{ i_1 i_2} C_{ j l k; i_1 i_2}^ {(d)} ( \vec \sigma( i_3), 
\vec p,\vec q,\vec q_0)
 \int  d^ 3 q_2t^ {(j)}_{ t_a T_a T_b} ( \vec p, \vec \pi( \vec q 
\vec q_2), E_q)
 G_0(\vec \pi'(\vec q \vec q_2) \vec q_2)\cr
& & t_{ t_c T_b T_0}^ { (l)}(\vec \pi' ( \vec q \vec q_2),\vec \pi 
( \vec q_2,\vec q_0),
E_{q_2})\phi_k( | \vec \pi'( \vec q_2,\vec q_0)|)
  \vec \sigma( i_1) \cdot \vec q_2 \vec \sigma( i_2) \cdot \vec q_2\cr
& + & C_{ j l k}^ {(e)}( \vec p,\vec q,\vec q_0)
  \int  d^ 3 q_2t^ {(j)}_{ t_a T_a T_b} ( \vec p, \vec \pi
( \vec q \vec q_2), E_q)
 G_0(\vec\pi'(\vec q \vec q_2) \vec q_2)\cr
& & t_{ t_cT_b T_0}^ { (l)}(\vec \pi' ( \vec q \vec q_2),\vec \pi 
( \vec q_2,\vec q_0),
E_{q_2})\phi_k( | \vec \pi'( \vec q_2,\vec q_0)|)
 \vec \sigma( i_1) \cdot \vec q_2 \vec \sigma( i_2)\cdot\vec q_2  
\sigma( i_3) \cdot \vec q_2 \cr
& + & \sum_{ i_1 \ne i_2} C_{ j l k; i_1 i_2}^ {(f)} ( \vec \sigma(
i_3) , 
\vec p,\vec q,\vec q_0)
 \int d^ 3 q_2 t^ {(j)}_{ t_a T_a T_b} ( \vec p, \vec \pi( \vec q 
\vec q_2), E_q)
 G_0( \vec \pi' ( \vec q \vec q_2) \vec q_2)\cr
& & t_{ t_c T_b T_0}^ { (l)}(\vec \pi' ( \vec q \vec q_2),\vec \pi 
( \vec q_2,\vec q_0),
E_{q_2})\phi_k( | \vec \pi'( \vec q_2,\vec q_0)|)  \vec \sigma( i_1 ) 
\times \vec \sigma( i_2) \cdot \vec q_2\cr
& + & \sum_{ i_1 \ne i_2 \ne i_3} 
C_{ jlk; i_1 i_2 i_3}^ {(g)} ( \vec p, \vec q, \vec q_0) \int d^ 3 q_2 t^
{(j)}_{ t_a T_a T_b} ( \vec p, \vec \pi( \vec q
\vec q_2), E_q)
 G_0( \vec \pi' ( \vec q \vec q_2) \vec q_2)\cr
& & t_{ t_c T_b T_0}^ { (l)}(\vec \pi' ( \vec q \vec q_2),\vec \pi
( \vec q_2,\vec q_0),
E_{q_2})\phi_k( | \vec \pi'( \vec q_2,\vec q_0)|)
 \vec \sigma(i_1) \cdot \vec q_2 ( \vec \sigma (i_2) 
\times \vec \sigma (i_3)) \cdot \vec q_2 \; \Big]\cr
& & P_{13}^ s P_{23}^ s | 1m_d\rangle | m_{10}\rangle  ~.
\label{e44}
\end{eqnarray}

\noindent
The integrals in Eq.~(\ref{e44}) are of the following form
\begin{eqnarray}
X_1&  = &  \int d^ 3 q_2 \; Y ~, \\
\label{e45}
X_2 & = &  \int d^ 3 q_2 \; Y \;\vec \sigma(i) \cdot \vec q_2 ~, \\
\label{e46}
X_3  & = &   \int d^ 3 q_2 \;  Y \; \vec \sigma(i) \cdot \vec q_2 
\vec \sigma(j) \cdot \vec q_2 ~, \\
\label{e47}
X_4  & = &   \int d^ 3 q_2\;  Y \;\vec \sigma(i) \cdot \vec q_2 
\vec \sigma(j) \cdot \vec q_2
\vec \sigma(k) \cdot \vec q_2 ~, \\
\label{e48}
X_5 & = &  \int d^ 3 q_2 \; Y \; \vec \sigma(i) \times \vec \sigma(j)  
\cdot \vec q_2 ~, \\
\label{e49}
X_6 & = &  \int d^ 3 q_2 \; Y \;\vec \sigma(i_1) 
\cdot \vec q_2 (\vec \sigma(i_2) \times \vec \sigma(i_3))
\cdot \vec q_2 ~,
\end{eqnarray}
where $Y$ represents functions with different dependencies on scalar quantities.
For instance in the first 
integral in Eq.~(\ref{e44}) on has
\begin{eqnarray}
Y  & = &    t^ {(j)}_{t_a T_a T_b}(\vec p,\vec \pi( \vec q \vec q_2), E_q)
\; G_0( \vec \pi' ( \vec q \vec q_2) \vec q_2) \cr
& &  t_{ t_c T_b T_0}^ { (l)}(\vec \pi' ( \vec q \vec q_2),\vec \pi (
 \vec q_2,\vec q_0),
E_{q_2}) \; \phi_k( | \vec \pi'( \vec q_2,\vec q_0)|) ~.
\label{e50}
\end{eqnarray}
As detailed in the Appendix C its angular dependence is
\begin{eqnarray}
Y = Y( \hat p \cdot \hat q, \hat p \cdot \hat q_2, \hat q \cdot \hat
q_2,  
\hat q \cdot \hat q_0,
\hat q_2 \cdot \hat q_0) ~.
\label{e51}
\end{eqnarray}

\noindent
The remaining integrals will be treated as exemplified for $X_2$. The integral
has the form
\begin{eqnarray}
X_2 & = &  \sigma_s(i)\int d^ 3 q_2 \; Y \;  q_{2s} ~.
\label{e52}
\end{eqnarray}
If $Y$ does not depend on $ \vec q_2 $ then $ X_2=0 $. Otherwise, due to
 the scalar nature of $Y$, one must have the following 
structure
\begin{eqnarray}
\int d^ 3 q_2 \; Y  \; q_{2,s} = \alpha p_s + \beta q_s + \gamma q_{0s} ~,
\label{e53}
\end{eqnarray}
and therefore
\begin{eqnarray}
X_2 & = & \alpha \vec \sigma(i) \cdot \vec p  + \beta\vec \sigma(i) 
\cdot \vec q + \gamma \vec \sigma(i)
\cdot \vec q_0 ~.
\label{e54}
\end{eqnarray}

\noindent
The scalars $ \alpha ,\beta,\gamma $ are then determined by multiplying
 Eq.~(\ref{e53}) by $ p_s$, $q_s$, and $q_{0s}$,
respectively. This leads to  three equations
\begin{eqnarray}
  \int d^ 3 q_2 Y  \vec p \cdot \vec q_2  & = & \alpha  p^ 2 +
  \beta \vec p \cdot \vec q + \gamma  \vec p \cdot \vec q_0 ~, \\
\label{e55}
 \int d^ 3 q_2 Y  \vec q \cdot \vec q_2   &= & \alpha  \vec q \cdot 
\vec p  + \beta q^ 2 + \gamma  \vec q \cdot \vec q_0 ~, \\
\label{e56}
 \int d^ 3 q_2 Y  \vec q_0 \cdot \vec q_2  & = & \alpha  \vec q_0
\cdot \vec p  + \beta \vec q_0 \cdot \vec q  + \gamma q_0^ 2 ~, 
\label{e57}
\end{eqnarray}
for  $ \alpha ,\beta,\gamma $. The three integrals  have to be
determined 
numerically.

\noindent
Correspondingly, despite being more involved, $X_3$ can be determined
\begin{eqnarray}
X_3 = \sigma_s(i) \sigma_t(j) \int d^ 3 q_2 Y q_{2s} q_{2t} ~.
\label{e58}
\end{eqnarray}
One has three external momenta in the function $ Y $, namely 
$ \vec p, \vec q $ , 
and $\vec q_0 $.
Thus, there are at most 9 possibilities: 
\begin{eqnarray}
 \int d^ 3 q_2 \; Y \; q_{2s} q_{2t} = \sum_{L=1}^ 9 A_L Q_{Ls} Q'_{ Lt} ~,
\label{e59}
\end{eqnarray}
where the $ Q_{Ls} Q'_{ Lt} $ belong to the set
\begin{eqnarray}
Q_{Ls} Q'_{ Lt} = [ p_s p_t, p_s q_t, p_s q_{0t}, q_s p_t, q_s q_t,
  q_s q_{0t}, 
q_{0s} p_t, q_{0s} q_t,
 q_{0s} q_{0t}] ~.
\label{e60}
\end{eqnarray}
By appropriate multiplications from the left one will have nine equations
for the coefficients $ A_L $,
generated by 9 corresponding integrals.
Thus $X_3$ takes the form
\begin{eqnarray}
X_3  = \sum_{L=1}^ 9 A_L \vec \sigma(i) \cdot \vec Q_L  \vec
\sigma(j) 
\cdot \vec Q'_L ~.
\label{e61}
\end{eqnarray}
Furthermore,
\begin{eqnarray}
X_4 = \sigma_s(i) \sigma_t(j) \sigma_u(k) \int d^ 3 q_2 \; Y \; q_{2s}
q_{2t} q_{2u} ~.
\label{e62}
\end{eqnarray}
Now we have 27 possibilities and consequently 27 equations need to be solved
 with the result
\begin{eqnarray}
X_4  = \sum_{L=1}^ {27} B_L \vec \sigma(i) \cdot \vec Q_L  \vec
\sigma(j) 
\cdot \vec Q'_L
\vec \sigma(k) \cdot \vec Q''_L ~,
\label{e63}
\end{eqnarray}
where $ Q_{Ls} Q'_{ Lt} Q''_{ Lu} $ are out of the set
\begin{eqnarray}
Q_{Ls} Q'_{ Lt} Q''_{ Lu} = [ p_s p_t p_u, p_s p_t q_u, p_s p_t q_{0u},\cdots ] ~.
\label{e64}
\end{eqnarray}
The appropriate 27 integrals need to be determined numerically. Finally,
$X_5$ can be handled like $X_2$, and $ X_6$ like $ X_3$.
All the integrals occurring in Eq.~(\ref{e44}) and the ones following from 
that  are of the type
\begin{eqnarray}
H & \equiv&  \int d^ 3 q_2 t^ {(j)}_{ t_a T_a T_b} ( \vec p, \vec \pi( 
\vec q \vec q_2), E_q) \;
 G_0( \vec \pi' ( \vec q \vec q_2) \vec q_2)\cr
& & t_{ t_c T_b T_0}^ { (l)}(\vec \pi' ( \vec q \vec q_2),\vec \pi 
( \vec q_2,\vec q_0), E_{q_2})
\phi_k( | \vec \; \pi'( \vec q_2,\vec q_0)|) f ~,
\label{e65}
\end{eqnarray}
where $ f $ consist of scalar products of the momenta 
$ \vec p,\vec q,\vec q_0 $, and $ \vec q_2 $.

We use the decomposition of the  NN t-matrix
from Eq.~(\ref{e30}) in the 
pd isospin space and the deuteron pole
structure from Eq.~(\ref{e33}). Then we obtain
\begin{eqnarray}
H & = &  \int d^ 3 q_2 G_0( \vec \pi' ( \vec q \vec q_2) \vec
q_2)\phi_k
( | \vec \pi'( \vec q_2,\vec
q_0)|)f \cr
& & \left(\delta_{ t_a 0} t^ {(j)}_{0 \frac{1}{2} \frac{1}{2}}( \vec p,\vec  
\pi( \vec q \vec q_2), E_q)
 + \delta_{ t_a1} t^ {(j)}_{1 T_a T_b}(\vec p, \vec  \pi( \vec q \vec
 q_2), E_q)\right) \cr
& & \left(  \delta_{ t_c 0} t^ {(l)}_{0 \frac{1}{2} \frac{1}{2}} (\vec \pi'
(\vec q \vec q_2),
\vec \pi ( \vec q_2,\vec q_0), E_{q_2})
 + \delta_{ t_c 1} t^ {(l)}_{1 T_b T_b}( \vec \pi' ( \vec q \vec q_2),
\vec \pi ( \vec q_2,\vec
q_0),E_{q_2} )\right) \cr
& = & \int d^ 3 q_2 G_0( \vec \pi' ( \vec q \vec q_2) \vec q_2) 
\phi_k( | \vec \pi'( \vec q_2,\vec q_0)|)  \cr
& & \left(\delta_{ t_a 0}
 \frac{ \hat t_{np}^ {(00,j)} (\vec p, \vec  \pi( \vec q \vec q_2), E_q)}
{ E_q + i \epsilon - E_d}
 + \delta_{ t_a 1} t^ {(j)}_{1 T_a T_b}(\vec p, \vec  \pi( \vec q \vec q_2), 
E_q) \right) \cr
& & \left(  \delta_{ t_c 0}
\frac{ \hat t_{np}^ {(00,l)} (\vec \pi' ( \vec q \vec q_2),\vec \pi 
( \vec q_2,\vec q_0), E_{q_2})}
{ E_{q_2} + i \epsilon - E_d}
  +   \delta_{ t_c 1} t^ {(l)}_{1 T_b T_b}( \vec \pi' 
( \vec q \vec q_2),\vec \pi ( \vec q_2,\vec q_0),
 E_{q_2}) \right) ~.
\label{e66}
\end{eqnarray}
In numerical implementation one can either follow 
 the standard path~\cite{report} 
 dealing with the free propagator singularity and 
 leading to moving logarithmic singularities, whose
treatment is well controlled and documented in~\cite{report,Liu:2004tv}  or 
one  applies the new way~\cite{new_method1,new_method2}
  which avoids that complication totally.
 We suggest  
the second path, which for the convenience of the
reader is detailed again in the Appendix D.

Summarizing, the second order amplitude of Eq.~(\ref{e42}) using all  
information given above will have the structure
\begin{eqnarray}
 \langle \vec p \vec q| \langle \gamma_a| t P G_0 t P | \Phi\rangle 
= \sum_f S_f( 
\vec p, \vec q, \vec q_0) 
 O_f( \vec \sigma(1), \vec \sigma(2), \vec \sigma(3), \vec p, \vec q,
 \vec q_0)P_f| 1 m_d\rangle | m_{10}\rangle  ~,
\label{e67}
\end{eqnarray}
where $O_f$ are spin-momentum dependent scalar operators and $ S_f $ 
only  momentum dependent
scalar functions, all depending just on the external momenta  $ \vec p, 
\vec q$,  and $\vec q_0 $. Further $ P_f$ is either 
$1, P_{12}^ s P_{23}^ s $ or $ P_{13}^ s P_{23}^ s $.

This form will be the input for the next order in Eq.~(\ref{e4}), namely 
$ t P G_0(t P G_0 t P | \Phi\rangle)$ where
the kernel $ t P G_0 $ is exactly treated as above and the first 
order term $ t P |\Phi\rangle $ is replaced by 
that second order form. Obviously this will go on like that and 
the total amplitude
  $ \langle \vec p \vec q | \langle \gamma_a | T | \Phi\rangle $ will 
appear in the 
form  of the right hand side of Eq.~(\ref{e67}).

It remains to present the calculation of the observables, which is 
described for the example of the breakup cross section in the next section.

\section{The nd breakup cross section}
\label{sec5}

The full breakup amplitude  according to Eq.~(\ref{e3}), projected on spin 
states  and using Eq.~(\ref{e10}) is given by
\begin{eqnarray}
\lefteqn{ \langle m_1| \langle m_2| \langle m_3|  \langle \vec p \vec
  q 
|\langle \gamma_a| (1 + P) T | 
\Phi\rangle } \cr
& = & \langle m_1| \langle m_2| \langle m_3|  \langle \vec p \vec q 
|\langle \gamma_a|  T | \Phi\rangle 
 +  \langle m_1| \langle m_2| \langle m_3|  \langle \vec p \vec q 
|\langle \gamma_a| P T | \Phi\rangle  \cr
 & = & \langle m_1| \langle m_2| \langle m_3|  \langle \vec p \vec q 
|\langle \gamma_a|  T | \Phi\rangle \cr
 & & +  \sum_b F_{ t_a t_b T_a} \delta_{ T_a T_b} \langle m_1| \langle
m_2| 
\langle m_3| 
  \langle \vec p \vec q | ( P_{12}^ {sm} P_{23}^ {sm} + (-)^ { t_a +
    t_b} 
P_{13}^ {sm} P_{23}^ {sm}) \langle \gamma_b|  T | \Phi\rangle ~.
\label{e68}
\end{eqnarray}
The simplest approach  is to apply the permutations to the left, using
\begin{eqnarray}
\langle \vec p \vec q| P_{13}^ m P_{23}^ m &  = & 
 \langle - \frac{1}{2} \vec p + \frac{3}{4} \vec q , - \vec  p -
 \frac{1}{2} 
\vec q| 
 \equiv \langle \vec p^ {~(1)} \vec q^ {~(1)}| ~, \\
\label{e69}
\langle \vec p \vec q| P_{12}^ m P_{23}^ m & = &   \langle - \frac{1}{2} 
\vec p - \frac{3}{4} \vec q , 
 \vec  p - \frac{1}{2} \vec q|  \equiv \langle \vec p^ {~(2)} \vec q^ {~(2)}|
 ~.
\label{e70}
\end{eqnarray}
Therefore,
\begin{eqnarray}
\lefteqn{\langle m_1| \langle m_2| \langle m_3|  \langle \vec p \vec q 
|\langle \gamma_a| (1 + P) T | 
\Phi\rangle  }\cr
& = & \langle m_1| \langle m_2| \langle m_3|  \langle \vec p \vec q 
|\langle \gamma_a|  T | \Phi\rangle\cr
& + & \sum_b F_{ t_a t_b T_a} \delta_{ T_a T_b}
     \langle m_1|   \langle m_2|   \langle m_3| P_{ 12}^ s P_{23}^ s 
   \langle {\vec p}^ {~(2)} {\vec q}^ {~(2)}| \langle \gamma_b | T 
| \Phi\rangle\cr
& + & \sum_b F_{ t_a t_b T_a} \delta_{ T_a T_b} (-)^ { t_a + t_b}
     \langle m_1|  \langle m_2| \langle m_3| P_{ 13}^ s P_{23}^ s  
 \langle {\vec p}^ {~(1)} {\vec q}^ {~(1)}|  \langle \gamma_b | T 
| \Phi\rangle ~.
\label{e71}
\end{eqnarray}

Using now the general form of the right hand side of Eq.~(\ref{e67}) one obtains
\begin{eqnarray}
\lefteqn{\langle m_1| \langle m_2| \langle m_3|  \langle \vec p \vec q 
|\langle \gamma_a| (1 + P) T |
 \Phi\rangle } \cr
& = & \langle m_1| \langle m_2| \langle m_3| \sum_f S_f ( \vec p, \vec
q, 
\vec q_0) O_f( 
\vec \sigma(1), \vec \sigma(2), \vec
\sigma(3), \vec p, \vec q,  \vec q_0) P_f | 1m_d\rangle | m_{10}\rangle \cr
& + & \sum_b F_{ t_a t_b T_a} \delta_{ T_a T_b}
     \langle m_1|   \langle m_2|   \langle m_3| P_{ 12}^ s P_{23}^ s\cr
& &  \sum_f S_f ( {\vec p}^ {~(2)}, {\vec q}^ {~(2)}, \vec q_0)  O_f( 
\vec \sigma(1), \vec \sigma(2), \vec
\sigma(3), {\vec p}^ {~(2)}, {\vec q}^ {~(2)}, \vec q_0 ) 
P_f | 1m_d\rangle | m_{10}\rangle\cr
& + & \sum_b F_{ t_a t_b T_a} \delta_{ T_a T_b} (-)^ { t_a + t_b}
     \langle m_1|  \langle m_2| \langle m_3| P_{ 13}^ s P_{23}^ s\cr
& & \sum_f S_f ( {\vec p}^ {~(1)}, {\vec q}^ {~(1)}, \vec q_0)  O_f( 
\vec \sigma(1), \vec \sigma(2), \vec
\sigma(3), {\vec p}^ {~(1)}, {\vec q}^ {~(1)} ,\vec q_0 )P_f |
1m_d\rangle 
| m_{10}\rangle\cr
& \equiv &   \sum_{\alpha} f_{\alpha} ( \vec p, \vec q, \vec q_0) 
 \langle m_1 m_2 m_3 | o_{\alpha}
( \vec \sigma(1), \vec \sigma(2),\vec \sigma(3), \vec p, \vec q, 
\vec q_0) P_{\alpha}| 1 m_d\rangle | m_{10}\rangle ~.
\label{e72}
\end{eqnarray}

As shown in Ref.~\cite{report}, the density matrix for the final state 
is given by
\begin{eqnarray}
( \rho_f)_{ij} = \sum_{kl} N_{ik} ( \rho_i)_{kl} N^ \dagger_{lj} ~,
\label{e73}
\end{eqnarray}
where 
\begin{eqnarray}
N_{ij} = \langle \Lambda_i | \langle \vec p \vec q | U_0 |\phi\rangle 
| \vec q_0\rangle | \lambda_j\rangle
~, \label{e76}
\end{eqnarray}
  is the expression of Eq.~(\ref{e72}),
 where the set of spin magnetic quantum numbers is given by
\begin{eqnarray}
{ | \Lambda_i\rangle } & =& { | m_1\rangle | m_2\rangle | m_3\rangle } ~,\cr
{ | \lambda_j\rangle }&= &{ | m_{10}\rangle | 1 m_d\rangle } ~,
\label{e74-75}
\end{eqnarray}
Furthermore,  $ \rho_i$ is the  density matrix for the initial state, 
\begin{eqnarray}
\rho_i = \frac{1}{6} Tr(\rho_i) \sum_{\nu} S^ { \nu} \langle S^ {\nu}\rangle_i ~,
\label{e77}
\end{eqnarray}
with $ [S^ {\nu} ]$ being the complete set of 2N spin matrices~\cite{report}.
For an unpolarized initial state one has
\begin{eqnarray}
\rho_i = \frac{1}{6} Tr(\rho_i) 1_{ 2\times 2} \times 1_{3\times 3} ~.
\label{e78}
\end{eqnarray}
With this, the unpolarized breakup cross section is given up to a 
phase-space factor as
\begin{eqnarray}
\sigma & \sim &  \frac{ Tr( \rho_f)}{ Tr ( \rho_i)}  =   \frac{1}{6} 
\sum_i \sum_k N_{ ik} N_{ik}^ *\cr
& = &  \frac{1}{6} \sum_i \sum_k | \langle \Lambda_i | 
\langle \vec p \vec q | U_0 |  \phi\rangle | \vec q_0\rangle 
| \lambda_k\rangle|^ 2\cr
& = &  \frac{1}{6} \sum_{ m_1 m_2 m_3} \sum_{ m_d m_{10}} \langle m_1 m_2 m_3 | 
\langle \vec p \vec q | U_0 | \phi\rangle | \vec q_0\rangle 
|  1 m_d\rangle | m_{10}\rangle\cr
& &  \langle  1 m_d| \langle m_{10}| \langle \vec q_0|\langle \phi |  
U_0^ \dagger | 
\vec p \vec q\rangle  | m_1 m_2 m_3\rangle ~.
\label{e79}
\end{eqnarray}
Using now the final form in Eq.~(\ref{e72}) we obtain
\begin{eqnarray}
\sigma & \sim &  \frac{1}{6} \sum_{ m_1 m_2 m_3} \sum_{ m_d m_{10}} 
\sum_{\alpha} f_{\alpha} ( \vec p, \vec q, \vec q_0)\cr
& & \langle m_1 m_2 m_3 | o_{\alpha} ( \vec \sigma(1), \vec \sigma(2),
\vec \sigma(3), \vec p, \vec q, 
\vec q_0) P_{\alpha}| 1 m_d\rangle | m_{10}\rangle\cr
& & \sum_{\beta} f_{\beta}^ * ( \vec p, \vec q, \vec q_0)
 \langle1  m_d | \langle m_{10}|P_{\beta}  o_{\beta}^ \dagger( \vec \sigma(1), 
\vec \sigma(2),\vec \sigma(3), \vec p, \vec q, \vec q_0) |m_1 m_2 m_3 \rangle\cr
&   = &  \frac{1}{6}\sum_{\alpha, \beta} f_{\alpha} f_{\beta}^ * 
\sum_{ m_1 m_2 m_3}
\langle m_1 m_2 m_3 | o_{\alpha} P_{\alpha} P_{\beta} o_{\beta}^ {\dagger} 
|  m_1 m_2 m_3\rangle ~.
\label{e80}
\end{eqnarray}

\section{Summary and Conclusions}
\label{sec6}

We extended the recently developed formalism 
for a new treatment of two- and three-nucleon bound states in three dimensions
to the realm of nucleon-deuteron   scattering. 
 The aim is to formulate the momentum space Faddeev equations
in such a fashion that the equations are reduced to 
 spin independent scalar 
functions in the same spirit as
already worked out for the 2N and 3N bound states. 
 We use the original, most general structure of NN forces, which is
 built 
out of scalar
operators of  spin- and momentum-vectors plus 
 scalar functions that only depend on momenta. This structure 
carries over to the NN  t-operator which is a central building block
in the Faddeev scheme. Generating the multiple 
scattering series for the Faddeev equations we arrive at the general
structure of the multiple scattering terms in the form 
of scalar spin-momentum dependent operators 
and momentum dependent scalar functions. 
Since our formulation only depends on operator and momentum vectors, there are
no intrinsic limitations on the energy range in which it can be applied. This is
contrast to the partial wave projected form of the momentum space
Faddeev 
equations.

Since three nucleon  forces appear naturally in the form of scalar
operators in  spin- and momentum-vectors and  
 scalar functions depending only on momenta, the present formulation
 can be 
extended to
include also 3N forces in a straightforward manner.
 This feature is especially important in view of the chiral
perturbation theory approach, where the multitude of 3N forces 
at N$^3$LO appears to be most
 efficiently treated directly in this three-dimensional formulation 
without having to expand these forces into  a partial wave basis.

\section*{Acknowledgments}
This work was supported by the Polish 2008-2011 science funds as the 
 research project No. N N202 077435 and in part under the
 auspices of the U.~S.  Department of Energy, Office of
 Nuclear Physics under contract No. DE-FG02-93ER40756
 with Ohio University.
 It was also partially supported by the Helmholtz
 Association through funds provided to the virtual institute ``Spin
 and strong QCD''(VH-VI-231) and by  
  the European Community-Research Infrastructure
Integrating Activity
``Study of Strongly Interacting Matter'' (acronym HadronPhysics2, 
Grant Agreement n. 227431)
under the Seventh Framework Programme of EU.

\appendix
\section{The scalar expressions $ a_{jk} $ and $ b_{jk}$}
\label{ap1}

Since the  scalar expressions $ a_{jk} $ and $ b_{jk}$ from
Eqs. (\ref{e25}) and (\ref{e26}) have to be determined only
once and their number is not too large, we provide all of them:
\begin{eqnarray}
a_{11}& = & 1\\
a_{12} &= & -  \frac{1}{3} \pi'^2 +  (\vec \sigma(1) \cdot \vec \pi')
( \vec \sigma(3) \cdot \vec \pi')\\
a_{21} &= &  \vec \sigma(2) \cdot \vec \sigma(3)\\
a_{22} &=& (\vec \sigma(1) \cdot \vec \pi') ( \vec \sigma(2) \cdot \vec \pi')\cr
&  - &  i (\vec \sigma(1) \cdot \vec \pi') (\vec \sigma(2) \times 
\vec \sigma(3)) \cdot \vec \pi'\cr
&  - &  \frac{1}{3} \pi'^2 (\vec \sigma(2) \cdot \vec \sigma(3))\\
a_{31} & = &  ( \vec \sigma(2) + \vec \sigma(3)) \cdot (\vec p \times 
\vec \pi)\\
a_{32} &=& \frac{3}{4} \vec q_0 \cdot (\vec p \times \vec q)  (\vec 
\sigma(1) \cdot \vec \pi')\cr
& - &  i (\vec \pi' \cdot \vec \pi) (\vec \sigma(1) \cdot \vec \pi')  
(\vec \sigma(3)) \cdot  \vec p)\cr
& + &  i (\vec p \cdot \vec \pi') (\vec \sigma(1) \cdot \vec \pi') 
(\vec \sigma(3) \cdot \vec \pi)\cr
& + &   (\vec \sigma(1) \cdot \vec \pi') (\vec \sigma(2)\cdot 
(\vec p \times \vec \pi)) ( \vec \sigma(3)
\cdot \vec \pi')\cr
&  - &   \frac{1}{3} \pi'^ 2 ( \vec \sigma(2) + \vec \sigma(3)) 
\cdot (\vec p \times \vec \pi)\\
a_{41} & = &  \vec \sigma(2) \cdot (\vec p \times \vec \pi)  
\vec \sigma(3) \cdot (\vec p \times \vec \pi)\\
a_{42} & = &  \frac{3}{4} (\vec p \times \vec q) \cdot \vec q_0  
(\vec \sigma(1) \cdot \vec \pi')
 \vec \sigma(2) \cdot (\vec p \times \vec \pi)\cr
& - &  i (\vec \pi' \cdot \vec \pi)  (\vec \sigma(1) \cdot 
\vec \pi') \vec \sigma(2) \cdot (\vec p \times
\vec \pi)  (\vec \sigma(3) \cdot \vec p)  \cr
& + &  i (\vec p \cdot \vec \pi')  (\vec \sigma(1) \cdot 
\vec \pi') \vec \sigma(2) \cdot (\vec p \times
\vec \pi) (\vec \sigma(3) \cdot \vec \pi) \cr
&  - &  \frac{1}{3} \pi'^ 2  \vec \sigma(2) \cdot (\vec p 
\times \vec \pi)  \vec \sigma(3) \cdot
(\vec p \times \vec \pi)\\
a_{51} & = & ( \vec \sigma(2) \cdot ( \vec p + \vec \pi)) 
( \vec \sigma(3) \cdot ( \vec p + \vec \pi))\\
a_{52} & = & ( ( \vec p + \vec \pi) \cdot \vec \pi') 
(\vec \sigma(1) \cdot \vec \pi')
 (\vec \sigma(2) \cdot ( \vec p + \vec \pi)) \cr
&  + &  i (\vec \sigma(1) \cdot \vec \pi') (\vec \sigma(2) 
\cdot ( \vec p + \vec \pi)) (\vec \sigma(3)
\cdot ( \vec p \times \vec \pi')) \cr
 & + & i \frac{3}{4} (\vec \sigma(1) \cdot \vec \pi') 
(\vec \sigma(2) \cdot ( \vec p + \vec \pi))
 (\vec \sigma(3) \cdot  (\vec q \times \vec q_0))\cr
&  - &  \frac{1}{3} \pi'^ 2 (\vec \sigma(2) \cdot 
( \vec p + \vec \pi)) (\vec \sigma(3) \cdot
 ( \vec p + \vec \pi))\\
a_{61} & = &  (\vec \sigma(2) \cdot ( \vec p - \vec \pi)) 
(\vec \sigma(3) \cdot ( \vec p - \vec \pi))\\
a_{62} & = &  (( \vec p - \vec \pi) \cdot \vec \pi')
( \vec \sigma(1) \cdot \vec \pi')
 (\vec \sigma(2) \cdot ( \vec p - \vec \pi)) \cr
&  + &  i (\vec \sigma(1) \cdot \vec \pi') (\vec \sigma(2) 
\cdot ( \vec p - \vec \pi)) (\vec \sigma(3)
\cdot ( \vec p \times \vec \pi'))\cr
&  - & i \frac{3}{4} (\vec \sigma(1) \cdot \vec \pi') 
(\vec \sigma(2) \cdot ( \vec p - \vec \pi)) (\vec
\sigma(3) \cdot (\vec q \times \vec q_0))\cr
&  - &  \frac{1}{3} \pi'^ 2 (\vec \sigma(2) \cdot ( \vec p - \vec \pi))
( \vec \sigma(3) \cdot ( \vec p - \vec \pi))
\end{eqnarray}

\begin{eqnarray}
b_{11}& = & 1\\
b_{12} &= &  - \frac{1}{3} \pi'^2 +  (\vec \sigma(1) 
\cdot \vec \pi') ( \vec \sigma(2) \cdot \vec \pi')\\
b_{21} &= &  \vec \sigma(2) \cdot \vec \sigma(3)\\
b_{22} &=& (\vec \sigma(1) \cdot \vec \pi') ( \vec \sigma(3) 
\cdot \vec \pi')\cr
&  + &  i (\vec \sigma(1) \cdot \vec \pi') (\vec \sigma(2) 
\times \vec \sigma(3)) \cdot \vec \pi'\cr
&  - &  \frac{1}{3} \pi'^2 (\vec \sigma(2) \cdot \vec \sigma(3))\\
b_{31} & = & - ( \vec \sigma(2) + \vec \sigma(3)) \cdot 
\vec p \times \vec \pi\\
b_{32} &=& -  \frac{3}{4} (\vec p \times \vec q) \cdot \vec q_0 
(\vec \sigma(1) \cdot \vec \pi')\cr
&  + &  i (\vec \pi \cdot \vec \pi') ( \vec \sigma(1) 
\cdot \vec \pi') ( \vec \sigma(2)\cdot \vec p)  \cr
&  - & i  (\vec p \cdot \vec \pi') (\vec \sigma(1) 
\cdot \vec \pi') (\vec \sigma(2) \cdot \vec \pi)\cr
&   - &  (\vec \sigma(1) \cdot \vec \pi') (\vec \sigma(2) 
\cdot \vec \pi') ( \vec \sigma(3) \cdot (\vec p \times
\vec \pi)) \cr
 & + &   \frac{1}{3} \pi'^ 2 ( \vec \sigma(2) +   
\vec \sigma(3)) \cdot (\vec p \times \vec \pi)\\
b_{41} & = &  \vec \sigma(2) \cdot (\vec p \times \vec \pi) 
\vec \sigma(3) \cdot (\vec p \times \vec \pi)\\
b_{42} & = &  \frac{3}{4} (\vec p \times \vec q) \cdot 
\vec q_0 (\vec \sigma(1) \cdot \vec \pi') \vec
\sigma(3) \cdot (\vec p \times \vec \pi)\cr
&  - &  i (\vec \pi \cdot \vec \pi') (\vec \sigma(1) 
\cdot \vec \pi') ( \vec \sigma(2) \cdot \vec p)
  \vec \sigma(3) \cdot (\vec p \times \vec \pi)\cr
& + &  i (\vec p \cdot \vec \pi') ( \vec \sigma(1) 
\cdot \vec \pi') (\vec \sigma(2) \cdot \vec \pi)
  \vec \sigma(3) \cdot (\vec p \times \vec \pi)\cr
&  - &  \frac{1}{3} \pi'^ 2 \vec \sigma(2) \cdot (\vec  p 
\times \vec \pi) \vec \sigma(3)
\cdot (\vec p \times \vec \pi)\\
b_{51} & = &  (\vec \sigma(2) \cdot ( \vec p - \vec \pi)) 
(\vec \sigma(3) \cdot ( \vec p - \vec \pi))\\
b_{52} & = &  (  \vec p - \vec \pi) \cdot \vec \pi' 
( \vec \sigma(1) \cdot \vec \pi')
 (\vec \sigma(3) \cdot ( \vec p - \vec \pi))\cr
&  + &  i (\vec \sigma(1) \cdot \vec \pi') (\vec \sigma(2) 
\cdot  (\vec p \times \vec \pi'))( \vec \sigma(3)
\cdot ( \vec p - \vec \pi))\cr
&  - & i  \frac{3}{4} (\vec \sigma(1) \cdot \vec \pi') 
(\vec \sigma(2)\cdot   (\vec q \times \vec q_0))
( \vec \sigma(3) \cdot ( \vec p - \vec \pi))\cr
&  - &  \frac{1}{3} \pi'^ 2 (\vec \sigma(2) \cdot ( \vec p 
- \vec \pi))( \vec \sigma(3) \cdot ( \vec p -
\vec \pi))\\
b_{61} & = &  (\vec \sigma(2) \cdot ( \vec p + \vec \pi)) 
(\vec \sigma(3) \cdot ( \vec p + \vec \pi))\\
b_{62} & = & ( (  \vec p + \vec \pi) \cdot \vec \pi') 
( \vec \sigma(1) \cdot \vec \pi') (\vec
\sigma(3) \cdot ( \vec p + \vec \pi))\cr
&  + &  i (\vec \sigma(1) \cdot \vec \pi') (\vec \sigma(2) 
\cdot  (\vec p \times \vec \pi')) (\vec \sigma(3)
\cdot ( \vec p + \vec \pi))\cr
&  + &  \frac{3}{4} i (\vec \sigma(1) \cdot \vec \pi') 
(\vec \sigma(2) \cdot  (\vec q \times \vec q_0))
 (\vec \sigma(3) \cdot ( \vec p + \vec \pi))\cr
&  - &  \frac{1}{3} \pi'^ 2 (\vec \sigma(2) \cdot ( \vec p + \vec \pi))
( \vec \sigma(3) \cdot ( \vec p + \vec \pi))
\end{eqnarray}

\section{Examples for the coefficients $ C_{jlk}$}
\label{ap2}

Again the coefficients $ C_{ jlk}, D_{jlk}, E_{jlk} $ and $ F_{ jlk}$ 
defined in Eqs.~(\ref{e38})-(\ref{e41}) have to be
calculated only once.  In the following we provide some examples,  
inserting   Eqs.~(\ref{e13}) for $ \vec \pi$
and $ \vec \pi'$ and removing the double occurrences of the same 
$ \vec \sigma(i)$:
\begin{eqnarray}
C_{111} & = & 1\\
C_{112} & = &  -  \frac{1}{3} ( \vec q_2 + \frac{1}{2} \vec q_0)^2\cr
& + & (\vec \sigma(2) \cdot \vec q_2)( \vec \sigma(1) \cdot \vec q_2)
 +(\vec \sigma(2) \cdot \vec q_2) ( \vec \sigma(1) \cdot \frac{1}{2} 
\vec q_0)\cr
&  + &  (\vec \sigma(2) \cdot \frac{1}{2} \vec q_0)( \vec \sigma(1) 
\cdot  \vec q_2)
+ (\vec \sigma(2) \cdot \frac{1}{2} \vec q_0) ( \vec \sigma(1) 
\cdot\frac{1}{2} \vec q_0)\\
C_{121} & = &  \vec \sigma(3) \cdot \vec \sigma(1)\\
C_{122}  & = &   (\vec \sigma(2) \cdot \vec q_2 )( \vec \sigma(3) 
\cdot\vec q_2)
 + (\vec \sigma(2) \cdot  \vec q_2) ( \vec \sigma(3) \cdot \frac{1}{2} 
\vec q_0)\cr
& + &  (\vec \sigma(2) \cdot \frac{1}{2} \vec q_0) ( \vec \sigma(3) 
\cdot \vec q_2) +
(\vec \sigma(2) \cdot\frac{1}{2} \vec q_0)( \vec \sigma(3) \cdot 
\frac{1}{2} \vec q_0)\cr
&  - &  i [ (\vec \sigma(2) \cdot \vec q_2) (\vec \sigma(3) \times 
\vec \sigma(1)) \cdot \vec q_2)
 + (\vec \sigma(2) \cdot \vec q_2) (\vec \sigma(3) \times \vec \sigma(1)) 
\cdot \frac{1}{2} \vec q_0)  \cr
&  + &  (\vec \sigma(2) \cdot\frac{1}{2} \vec q_0)(\vec \sigma(3) 
\times \vec \sigma(1)) \cdot   \vec q_2
)
 + (\vec \sigma(2) \cdot \frac{1}{2} \vec q_0)(\vec \sigma(3) 
\times \vec \sigma(1)) \cdot
\frac{1}{2} \vec q_0) ] \cr
&  - &  \frac{1}{3} ( \vec q_2 + \frac{1}{2} \vec q_0)^2  
(\vec \sigma(3) \cdot \vec \sigma(1))\\
C_{211}  & = &  \vec \sigma(2) \cdot \vec \sigma(3)\\
C_{212} & = & - \frac{1}{3} ( \vec q_2 + \frac{1}{2} \vec
q_0)^2 (\sigma(2) \cdot \vec \sigma(3)) \cr
& + &   (\vec \sigma(1) \cdot \vec q_2) ( \vec \sigma(3) \cdot \vec q_2)
 + (\vec \sigma(1) \cdot \vec q_2) ( \vec \sigma(3) \cdot \frac{1}{2} 
\vec q_0)\cr
&  + &  (\vec \sigma(1) \cdot\frac{1}{2} \vec q_0)( \vec \sigma(3) 
\cdot \vec q_2)
 + (\vec \sigma(1) \cdot\frac{1}{2} \vec q_0)( \vec \sigma(3) \cdot
\frac{1}{2} \vec q_0) \cr
&  - &  i [ (\vec \sigma(1) \cdot \vec q_2) (\vec \sigma(3) \times 
\vec \sigma(2)) \cdot \vec q_2)
 + (\vec \sigma(1) \cdot \vec q_2) (\vec \sigma(3) \times \vec \sigma(2)) 
\cdot \frac{1}{2} \vec q_0)  \cr
&  + &  (\vec \sigma(1) \cdot\frac{1}{2} \vec q_0)(\vec \sigma(3) 
\times \vec \sigma(2)) \cdot   \vec q_2
)
 + (\vec \sigma(1) \cdot \frac{1}{2} \vec q_0)(\vec \sigma(3) 
\times \vec \sigma(2)) \cdot
\frac{1}{2} \vec q_0) ] \\
C_{221} & = &\vec \sigma(1) \cdot \vec \sigma(2) - i \vec \sigma(3) 
\cdot \vec \sigma(1) \times \vec
\sigma(2)\\
C_{222} & = &  -i (\vec \sigma(3) \cdot  ( \vec q_2 + \frac{1}{2} \vec q_0)
(\vec \sigma(2) \times \vec \sigma(1)) \cdot  (\vec q_2 + \frac{1}{2}
\vec q_0)\cr
&  - &  \frac{1}{3} ( \vec q_2 + \frac{1}{2} \vec q_0)^2 
[ (\vec \sigma(1) \cdot \vec \sigma(2)) 
- i \vec \sigma(3) \cdot  (\vec \sigma(1) \times \vec \sigma(2)) ] \cr
& + & ( \vec q_2 + \frac{1}{2} \vec q_0)^2 [ 1 + 
 \vec \sigma(1) \cdot  \vec \sigma(3) -  \vec \sigma(1) \cdot  \vec
 \sigma(2) - \vec \sigma(2) \cdot  \vec \sigma(3) ] \cr 
& + &  ( \vec \sigma(1) \cdot  ( \vec q_2 + \frac{1}{2} \vec q_0)) 
        ( \vec \sigma(2) \cdot  ( \vec q_2 + \frac{1}{2} \vec q_0)) 
       -( \vec \sigma(1) \cdot  ( \vec q_2 + \frac{1}{2} \vec q_0)) 
        ( \vec \sigma(3) \cdot  ( \vec q_2 + \frac{1}{2} \vec q_0)) \cr
& + &   ( \vec \sigma(2) \cdot  ( \vec q_2 + \frac{1}{2} \vec q_0)) 
        ( \vec \sigma(3) \cdot  ( \vec q_2 + \frac{1}{2} \vec q_0))  
\end{eqnarray}

\section{The angular dependence of Y}
\label{ap3}

In order to see the angular dependence of a term like that given 
in (\ref{e50}) we use
\begin{eqnarray}
\vec \pi( \vec q \vec q_2) & = & \frac{1}{2} \vec q + \vec q_2\\
\vec \pi'( \vec q_2 \vec q_0) & = & - \vec q_2 - \frac{1}{2}  \vec q_0\\
\vec \pi( \vec q_2 \vec q_0) & = &  \frac{1}{2} \vec q_2 + \vec q_0\\
\vec \pi'( \vec q \vec q_2) & = & - \vec q - \frac{1}{2} \vec q_2
\end{eqnarray}
Therefore
\begin{eqnarray}
\hat p \cdot \hat \pi ( \vec q \vec q_2)&  = &  \frac{ \frac{1}{2} 
\hat p \cdot \vec q
+ \hat p \cdot \vec q_2}{ | \frac{1}{2} \vec q + \vec q_2|}\\
| \pi'( \vec q \vec q_2)| & = &  \sqrt { q_2^ 2 +  \frac{1}{4} q_0^ 2 
+ \vec q_2 \cdot \vec q_0}\\
\hat \pi'( \vec q \vec q_2) \cdot \hat \pi( \vec q_2 \vec q_0)&  = &
 - \frac{  \frac{1}{2} \vec q \cdot \vec q_2 + \vec q \cdot \vec q_0
 + \frac{1}{4} q_2^ 2 + \frac{1}{2} \vec q_2 \cdot \vec q_0}
{ | \vec q + \frac{1}{2} \vec q_2| | \frac{1}{2} \vec q_2 + \vec q_0|}
\end{eqnarray}
We put $ \hat q = \hat z $ and define
\begin{eqnarray}
x & = &  \hat q_2 \cdot \hat q\\
x_p & = &  \hat p \cdot \hat q\\
\hat p \cdot \hat q_2 & = &  x_p x + \sqrt{ 1 - x_p^ 2} 
\sqrt{ 1 - x^  2} cos ( \phi_p - \phi_2)\\
x_{ q_0} & = &  \hat q_0 \cdot \hat q\\
\hat q_2 \cdot \hat q_0 & = &  x x_{ q_0} + \sqrt{ 1 - x^ 2} 
\sqrt{ 1 - x_{ q_0}^ 2} cos ( \phi_2 - \phi_{ q_0})
\end{eqnarray}
This settles the angular dependencies of $ Y $.

\section{Treatment of singularities}
\label{ap4}

Regarding the expression (\ref{e66}) we face two types of integrals, 
where $ G_0 $ appears alone or together with
the deuteron pole:
\begin{eqnarray}
H_1 & \equiv &  \int d^ 3 q_2 g( \vec p, \vec \pi ( \vec q \vec q_2), 
\vec \pi' ( \vec q \vec q_2),
 \vec \pi ( \vec q_2 \vec q_0),  \vec \pi' ( \vec q_2 \vec q_0))  
G_0( \vec \pi' ( \vec q \vec q_2) \vec
q_2)\\
H_2 & \equiv &  \int d^ 3 q_2  h ( \vec p, \vec \pi ( \vec q \vec
q_2),
\vec \pi' ( \vec q \vec q_2),
 \vec \pi ( \vec q_2 \vec q_0),\vec \pi' ( \vec q_2 \vec q_0))\cr
& & G_0( \vec \pi' ( \vec q \vec q_2) \vec q_2)  \frac{1} { E_{q_2} 
+ i \epsilon - E_d}
\end{eqnarray}
Here $g$ and $h$ are regular scalar functions.
 In order to arrive at the new way~\cite{new_method2} one has to go
 back 
and rewrite both integrals into
\begin{eqnarray}
H_1 & \equiv &  \int d^ 3 q_2  \int d^ 3 p' \int d^ 3 p_2 \delta( \vec
p~' 
- \vec \pi ( \vec q \vec q_2))
\delta( \vec p_2 - \vec \pi' ( \vec q \vec q_2))\cr
& &  g( \vec p, \vec p', \vec p_2, \vec \pi ( \vec q_2 \vec q_0),\vec
\pi' 
( \vec q_2 \vec q_0))
 G_0( \vec p_2, q_2)\\
H_2 & \equiv &  \int d^ 3 q_2  \int d^ 3 p' \int d^ 3 p_2 \delta( \vec
p~' 
- \vec \pi ( \vec q \vec q_2))
\delta( \vec p_2 - \vec \pi' ( \vec q \vec q_2)) h ( \vec p, \vec p~', \vec p_2,
\vec \pi ( \vec q_2 \vec q_0),\vec \pi' ( \vec q_2 \vec q_0))\cr
& & G_0( \vec p_2,  \vec q_2)  \frac{1} { E_{q_2} + i \epsilon - E_d}
\end{eqnarray}
Then we change the product of the following two $ \delta $-functions
\begin{eqnarray}
& & \delta( p' - \pi( \vec q \vec q_2)) \delta( p_2 
- \pi' ( \vec q \vec q_2)) \cr
 &=& \delta( p' - \sqrt{ \frac{1}{4} q^ 2 + q_2^ 2 + q q_2 x}) \delta(
  p_2 
- \sqrt{ q^ 2 + \frac{1}{4}
q_2^ 2  + q q_2 x})\cr
& = &    \frac{2 p'}{q q_2} \delta( x - x_0) \Theta( 1 - | x_0|)\cr
& &  \delta( p_2 - \sqrt{ \frac{3}{4} q^ 2 - \frac{3}{4} q_2^ 2  +
  {p'}^ 2})
\Theta( \frac{3}{4} q^ 2 -
\frac{3}{4} q_2^ 2  + {p'}^ 2) ~,
\end{eqnarray}
with
\begin{eqnarray}
x_0 = \frac{ {p'}^ 2 - \frac{1}{4} q^ 2 - q_2^ 2 }{ q q_2}  ~.
\end{eqnarray}
We start with $ H_1 $ and insert (D5)
\begin{eqnarray}
  H_1 & = &    \int d^ 3 q_2  \int d \hat p' \int d \hat p_2
\int dp' dp_2  \frac{2 p'}{q q_2} \delta( x - x_0) \Theta( 1 - | x_0|)\cr
& & \delta( p_2 - \sqrt{ \frac{3}{4} q^ 2 - \frac{3}{4} q_2^ 2  +
  {p'}^ 2})
\Theta( \frac{3}{4} q^ 2 -
\frac{3}{4} q_2^ 2  + {p'}^ 2) g( \vec p, \vec p~', \vec p_2, \vec \pi 
( \vec q_2 \vec q_0),
\vec \pi' ( \vec q_2 \vec q_0))\cr
& &  \frac{1}{ E + i \epsilon - \frac{3}{4m}q_2^ 2 - \frac{ p_2^ 2}{m}}
 \delta( \hat p' - \hat \pi( \vec q \vec q_2)) \delta( \hat p_2 
- \hat \pi'( \vec q \vec q_2)) ~.
 \end{eqnarray}
Then  we carry out the $ p_2, \hat p_2 $ and $ \hat p'$  integrations
\begin{eqnarray}
 H_1 & = &  \frac{2}{q}   \int d \hat  q_2 \int dp' p' dq_2 q_2  
\delta( x - x_0) \Theta( 1 - | x_0|)\cr
& & \Theta( \frac{3}{4} q^ 2 -\frac{3}{4} q_2^ 2  + {p'}^ 2)
g( \vec p, p' \hat \pi ( \vec q \vec q_2),  p_2 
\hat \pi'( \vec q \vec q_2), \vec \pi ( \vec q_2 \vec
q_0),
 \vec \pi' ( \vec q_2 \vec q_0))\cr
& &  \frac{1}{ E + i \epsilon - \frac{3}{4m}q_2^ 2 - \frac{1}{m} 
( \frac{3}{4} q^ 2 - \frac{3}{4} q_2^ 2
+ {p'}^ 2)}\cr
& = & \frac{2}{q}   \int dp' p' \int dq_2 q_2   \Theta( 1 - | x_0|)
\Theta( \frac{3}{4} q^ 2 -\frac{3}{4} q_2^ 2  + {p'}^ 2) \cr
& &  \tilde g(  p,  p', q,  q_2) \frac{1}{ E + i \epsilon -
  \frac{3}{4m}q ^ 2 
- \frac{1}{m} {p'}^ 2} ~,
 \end{eqnarray}
with
\begin{eqnarray}
p_2 = \sqrt{ \frac{3}{4}( q^ 2 - q_2 ^2) + {p'}^ 2} ~,
\end{eqnarray}
and
\begin{eqnarray}
& & \tilde g(  p,  p',q, q_2)\cr
&  = &  \int d \hat q_2 \delta( x - x_0)
g( \vec p, p' \hat \pi ( \vec q \vec q_2), p_2 \hat \pi'( \vec q \vec
q_2), 
\vec \pi ( \vec q_2 \vec
q_0),
\vec \pi' ( \vec q_2 \vec q_0)) ~.
\end{eqnarray}

Remember $ x = \hat q_2 \cdot \hat q $, thus it appears natural to
  choose 
$ \hat q = \hat z $ and then
the $ x $-integration can be carried out trivially.

The two $ \Theta$-functions restrict the integrations 
in $ q_2 $ and $ p'$ into an area whose size
depends on the magnitudes of the spectator momentum $ q $ . 
It results
\begin{eqnarray}
 H_1 & = & \frac{2}{q}\int_0^ { \infty}  dp' p'\frac{1}{ E + i
   \epsilon 
- \frac{3}{4m}q ^ 2 - \frac{1}{m}
{p'}^ 2}\cr
& &   \int_{| \frac{q}{2} - p'|}^ { \frac{q}{2} + p'} dq_2 q_2   
\tilde g(  p,  p',q,  q_2) ~.
 \end{eqnarray}
The important point is that the free propagator appears now as a simple pole.

Next comes the $ H_2 $-integral, where two singular denominators appear.
We rewrite it using again (D5) and obtain
\begin{eqnarray}
H_2 & = &    \int d^ 3 q_2  \int d \hat p' \int d \hat p_2
\int dp' dp_2  \frac{2 p'}{q q_2} \delta( x - x_0) \Theta( 1 - | x_0|)\cr
& & \delta( p_2 - \sqrt{ \frac{3}{4} q^ 2 - \frac{3}{4} q_2^ 2  
+ {p'}^ 2})\Theta( \frac{3}{4} q^ 2 -
\frac{3}{4} q_2^ 2  + {p'}^ 2) h( \vec p, \vec p~', \vec p_2, \vec \pi 
( \vec q_2 \vec q_0),
\vec \pi' ( \vec q_2 \vec q_0))\cr
& &  \frac{1}{ E + i \epsilon - \frac{3}{4m}q_2^ 2 - \frac{ p_2^
    2}{m}} 
\frac{1}{ E_{q_2} + i \epsilon - E_d} \delta( \hat p' 
- \hat \pi( \vec q \vec q_2)) \delta( \hat p_2 -
\hat \pi'( \vec q \vec q_2)) ~.
 \end{eqnarray}
Then we carry out again the $ p_2, \hat p_2 $ and $ \hat p'$ integrations
\begin{eqnarray}
H_2 & = &  \frac{2}{q}  \int d \hat q_2  \int dp'p' dq_2 q_2   
\delta( x - x_0) \Theta( 1 - | x_0|)\cr
& & \Theta( \frac{3}{4} q^ 2 - \frac{3}{4} q_2^ 2  + {p'}^ 2)
\tilde  h( p,  p',q, q_2)\cr
& &  \frac{1}{ E + i \epsilon - \frac{3}{4m}q_2^ 2 - \frac{ p_2^ 2}{m}}
 \frac{1}{ E_{q_2} + i \epsilon - E_d} ~,
 \end{eqnarray}
with
\begin{eqnarray}
& & \tilde h(  p,  p',q, q_2)\cr
&  = &  \int d \hat q_2 \delta( x - x_0)
h( \vec p, p' \hat \pi ( \vec q \vec q_2), p_2 \hat \pi'( \vec q \vec q_2),
 \vec \pi ( \vec q_2 \vec q_0), \vec \pi' ( \vec q_2 \vec q_0)) ~.
\end{eqnarray}
We rewrite
\begin{eqnarray}
& & \frac{1}{ E + i \epsilon - \frac{3}{4m}q ^ 2 - \frac{ {p'}^ 2}{m}}
\frac{1}{ E + i \epsilon - \frac{3}{4m}q_2^ 2 - E_d}\cr
&  = &  [ \frac{1}{ E + i \epsilon - \frac{3}{4m}q ^ 2 -
\frac{ {p'}^ 2}{m}} - \frac{1}{ E + i \epsilon - \frac{3}{4m}q_2^ 2 - E_d} ]\cr
& &  \frac{ 1 }{ - E_d - \frac{3}{4m}q_2^ 2 + \frac{1}{m} ( {p'}^ 2 
+ \frac{3}{4}q ^ 2)} ~.
\end{eqnarray}
The new denominator function
\begin{eqnarray}
\tilde G( q, q_2,p') \equiv \frac{ 1 }{ - E_d - \frac{3}{4m}q_2^ 2 
+ \frac{1}{m} ( {p'}^ 2
+ \frac{3}{4}q ^ 2)}
\end{eqnarray}
cannot become singular inside the integration domain $ p'-q_2 $. 
Using the expression (D9) we see
that
\begin{eqnarray}
\tilde G( q, q_2,p') = \frac{1}{ - E_d + \frac{1}{m} p_2^ 2 } 
= \frac{1}{ |E_d| + \frac{1}{m} p_2^ 2 } ~.
\end{eqnarray}
Therefore (D13) turns into
\begin{eqnarray}
H_2 & = &  \frac{2}{q}  \int d \hat q_2  \int dp'p' dq_2 q_2   
\delta( x - x_0) \Theta( 1 - | x_0|)\cr
& & \Theta( \frac{3}{4} q^ 2 - \frac{3}{4} q_2^ 2  + {p'}^ 2)
\tilde  h( p,  p',q, q_2)\cr
& & [ \frac{1}{ E + i \epsilon - \frac{3}{4m}q ^ 2 - \frac{ {p'}^ 2}{m}}
 - \frac{1}{ E + i \epsilon - \frac{3}{4m}q_2^ 2 - E_d} ] \tilde G( q, q_2,p')
\end{eqnarray}
We use the integration over the area in the $ p'-q_2 $ plane and obtain
\begin{eqnarray}
H_2 & = &  \frac{2}{q} \int_0^ {\infty} dp' p'
\frac{1}{ E + i \epsilon - \frac{3}{4m}q ^ 2 - \frac{ {p'}^ 2}{m}} 
\tilde  h( p,  p',q, q_2)
  \int_{| \frac{q}{2} - p'|}^ {\frac{q}{2} + p'} dq_2 q_2 \tilde G( q, q_2,p')\cr
& - & \frac{2}{q}  \int_0^ { \infty} dq_2 q_2  \frac{1}{ E 
+ i \epsilon - \frac{3}{4m}q_2^ 2 - E_d}
\tilde h( p,  p',q, q_2)
  \int_{ | \frac{q}{2} - q_2|}^ { \frac{q}{2} + q_2} dp' p'
  \tilde G( q, q_2,p')  ~.
\end{eqnarray}

In the second integral we
integrated first over $ q_2 $ and then over $ p' $.
In both cases we have just a simple pole either in $p'$ or in $q_2$.

\end{document}